\documentclass[
prl,
twocolumn,
showpacs,
groupedaddress,
superscriptaddress,
]{revtex4-1}

\usepackage{soul}
\usepackage{amsfonts,amstext}
\usepackage{graphicx}
\usepackage{dcolumn}
\usepackage{bm}
\usepackage[flushleft]{threeparttable}
\usepackage{url}
\usepackage{amsmath}
\usepackage{amssymb}
\usepackage{braket}
\usepackage{multirow}
\usepackage{array}
\usepackage{url}
\usepackage{booktabs}
\usepackage{float}
\usepackage{lineno,hyperref}
\usepackage{mathtools}
\usepackage{xcolor}
\usepackage{MnSymbol}
\usepackage{enumitem}
\usepackage{setspace}

\hypersetup{colorlinks=true, urlcolor=blue, citecolor=cyan, pdfborder={0 0 0}}
\newcommand{\vv}[1]{{\boldsymbol #1}}

\newcolumntype{P}[1]{>{\centering\arraybackslash}p{#1}}

\newcommand{\cyan}[1]{{\em{\color{cyan}{{#1}}}}}

\newcommand{\nn}{\nonumber \\}

\makeatletter
\newcommand{\newparallel}{\mathrel{\mathpalette\new@parallel\relax}}
\newcommand{\new@parallel}[2]{%
  \begingroup
  \sbox\z@{$#1T$}
  \resizebox{!}{\ht\z@}{\raisebox{\depth}{$\m@th#1/\mkern-5mu/$}}%
  \endgroup
}
\makeatother

\begin{document}

\setstcolor{blue} 
\title{Two-dimensional weak-type insulators in inversion-symmetric crystals}
\author{Sunam Jeon}
\affiliation{Department of Energy Science, Sungkyunkwan University, Suwon 16419, Korea}
\author{Youngkuk Kim}
\email{youngkuk@skku.edu}
\affiliation{Department of Physics, Sungkyunkwan University, Suwon 16419, Korea}
\date{\today}

\begin{abstract} 
The Su-Schrieffer-Heeger (SSH) chain is an one-dimensional lattice that comprises two dimerized sublattices. Recently, Zhu, Prodan, and Ahn (ZPA) proposed in [L. Zhu, E. Prodan, and K. H. Ahn, Phys. Rev. B \textbf{99}, 041117 (2019)] that one-dimensional flat bands can occur at topological domain walls of a two-dimensional array of the SSH chains. Here, we newly suggest a two-dimensional topological insulator that is protected by inversion and time-reversal symmetries without spin-orbit coupling. It is shown that the two-dimensional SSH chains realize the proposed topological insulator. Utilizing the first Stiefel-Whitney numbers, a weak type of $\mathbb{Z}_2$ topological indices are developed, which identify the proposed topological insulator, dubbed a two-dimensional Stiefel-Whitney insulator (2DSWI). The ZPA model is employed to study the topological phase diagrams and topological phase transitions. It is found that the phase transition occurs via the formation of the massless Dirac points that wind the entire Brillouin zone. We argue that this unconventional topological phase transition is a characteristic feature of the 2DSWI, manifested as the one-dimensional domain wall states. Using first-principles calculations, we find the suggested 2DSWI should be realized in eleven known materials, such as Zn$_2$(PS$_3$)$_3$. The new insight from our work could help efforts to realize topological flat bands in solid-state systems.
\end{abstract}
\maketitle

\cyan{Introduction -} Recently, there has been a surge of interest in the correlated electrons in flat bands \cite{Wu07p070401,Wong13p060504,Biesenthal19p183601,Balents20p725}. In particular, twisted bilayer graphene \cite{Santos07p256802,Dean10p722,Morell10p121407,Bistritzer11p12233,Wang13p614,Kim16p1989}, in which interlayer twist angle tunes the bandwidth to zero, has spurred a great deal of community interest, exhibiting strong correlations of electrons and nontrivial band topology \cite{Cao18p43,Cao18p80,Song19p036401,Brihuega19p196802}. The advent of magic angle twisted bilayer graphene has renewed the interest in the realization of flat bands in diverse systems \cite{Heikkil011p233, Fu19p035448, hwang21p2105.14919}. Notable examples include the recent study by Zhu, Prodan, and Ahn (ZPA) \cite{Zhu19p041117}. Utilizing the two-dimensional (2D) array of the Su-Schrieffer-Heeger (SSH) chains \cite{SSH79p1698,PhysRevB.22.2099,RevModPhys.60.781}, ZPA have shown that one-dimensional (1D) flat bands can arise, being stabilized via the formation of topological domain walls (DWs). Encouragingly, the proposed 1D flat bands have been experimentally confirmed in a mechanical metamaterial \cite{Qian20p225501}, whereas their condensed-matter realization is yet to be discovered.

Regarding efforts to search for topological materials, these days have witnessed remarkable developments, based on topological quantum chemistry and symmetry indicators \cite{Slager2012, PhysRevB.86.115112, PhysRevX.7.041069, Bradlyn2017, Po17p1, PhysRevX.8.031070, PhysRevB.98.115150, PhysRevB.100.195135, Tang19peaau8725, Tang19p470}. Due to these high-throughput approaches, mass topological materials have been catalogued \cite{Tang19p486, Zhang19p475,Vergniory2019,Maia21p2105.09954}, started with the seminal work by Fu and Kane \cite{Fu07p045302}. The Fu-Kane formula expresses the $\mathbb{Z}_2$ topological indices using parity eigenvalues, enabling the discovery of archetypal topological materials, such as  Bi$_{1-x}$Sb$_x$ \cite{Hasan11p55, Hsieh08p970} and Bi$_2$Se$_3$ \cite{Xia09p398, Zhang09p053114}. Later, the strong $\mathbb{Z}_2$ index has been extended to the $\mathbb{Z}_4$ index $\nu_0$, including a wider class of topological phases with and without spin-orbit coupling (SOC) \cite{Benalcazar2017, PhysRevX.8.031070, PhysRevLett.119.246401, PhysRevLett.119.246402, Schindler2018, PhysRevX.10.031001}. For example, without SOC, $\nu_0$ = 2 characterizes a $\mathbb{Z}_2$ monopole nodal-line semimetal in three dimensions \cite{Ahn18p106403, Zhijun19p186401}, whereas it characterizes higher-order topological insulators in two dimensions \cite{Park19p216803, Lee20p1}. Interestingly, an insulators with $\nu_0 = 0$ can contain nontrivial band topology, similar to inversion-symmetric topological insulators \cite{Hughes11p245132, Alexandradinata14p155114}. However, this opportunity to encounter a novel topological phase has remained widely unaddressed to date. 

In this Letter we propose and characterize a novel topological class of insulators with $\nu_0 = 0$ under inversion $\mathcal{P}$ and time-reversal $\mathcal{T}$ symmetries without SOC, dubbed the 2D Stiefel-Whitney insulator (2DSWI). Utilizing the first Stiefel-Whitney (SW) numbers, we develop two weak $\mathbb{Z}_2$ indices ($\nu_1\nu_2$), which complement the strong $\mathbb{Z}_2$ index $\nu_0 = 0$ and identify the 2DSWIs. Employing the ZPA model, we show that the 2DSWI is realized in the 2D coupled SSH chains under various centrosymmetric deformations. The topological phase diagrams are found to facilitate four distinct 2DSWI phases, indexed by $(\nu_0;\nu_1\nu_2) = (0;00),(0;01),(0;10)$, and $(0;11)$, respectively. In between distinct 2DSWIs, a topological semimetal with $\nu_0 = 1$ appears, featuring a pair of gapless Dirac points (DPs). We show that the DPs mediate a topological phase transition by winding the Brillouin zone (BZ), and their trajectories manifest as 1D DW states between distinct 2DSWIs.

\cyan{$\mathbb{Z}_2$ topological indices -} We consider a 2D Bloch Hamiltonian $\mathcal{H}(\vv k)$ that is invariant under $\mathcal {PT}$ and $\mathcal{T}^2 = 1$. Since $\left[\mathcal{PT},\mathcal H (\vv k)\right] = 0$, 1D families of real Hamiltonians on a closed loop $\mathcal C$ in the BZ, $\mathcal{H}(\vv k)|_{\vv k \in \mathcal C}$, can be characterized by the $\mathbb{Z}_2$-quantized first SW number $\nu_{\mathcal{C}} = \frac{1}{\pi} P \oint_{ \mathcal{C}} \vv A(\vv k)\cdot d\vv k$ \cite{Ahn19p117101}, where $P$ is the path-ordering operator, $\vv A (\vv k) = i \langle u(\vv k)|\vv \nabla_{\vv k} |u(\vv k)\rangle$ is the Berry connection, and $u(\vv k)$ is the cell-periodic part of the occupied Bloch function at $\vv k \in$ BZ. The  number refers to a topological invariant that characterizes the twist of real Bloch states in momentum space \cite{Ahn19p117101}. The Zak phase \cite{Zak89p2747} that is $\mathbb{Z}_2$-quantized by the reality condition corresponds to the first SW number. The SW number is defined for any given $k_1$ and $k_2$ lines as $\nu_1 (k_1) = \frac{1}{\pi}P\int_{-\pi}^{\pi} A_1 (k_1, k_2) dk_2$ and $\nu_2 (k_2) = \frac{1}{\pi}P\int_{-\pi}^{\pi} A_2 (k_1, k_2) dk_1$, where $A_i$ ($i$ = 1, 2) is $k _i$-component of $\vv A(\vv k)$. In the presence of a direct band gap, one-parameter families of real Hamiltonians on a $k_i$ line ($i = 1, 2$) are adiabatically connected to those on a $k'_i$ line for any $k'_i \ne k'_i$, resulting in the equivalent SW numbers $\nu_i(k_i) \equiv \nu_i(k'_i)$. Therefore, the $\mathbb{Z}_2$ indices $\nu_1$ and $\nu_2$ can unambiguously characterize a $\mathcal{PT}$-symmetric insulating phase.

\begin{figure}
\includegraphics[width=0.45\textwidth]{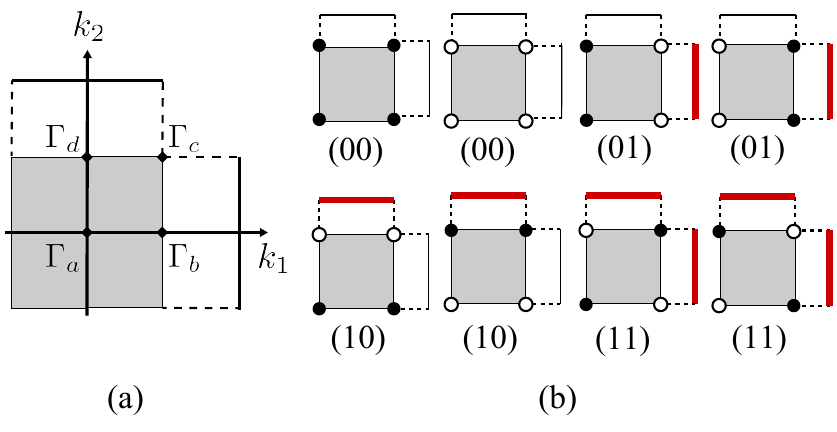}%
\caption{Diagrams depicting possible 2DSWIs indexed by ($\nu_1\nu_2$) (a) 2D bulk BZ and 1D edge BZ. $\Gamma_i$ are TRIMs. (b) TRIMs with $\xi_i = +1(-1)$ are denoted by open (closed) circles at $\Gamma_i$. The thick (red) line indicates possible edge states on the $(\nu_1\nu_2)$ edge.
} 
\label{fig:sw_numbers}
\end{figure}

\cyan{Inversion-symmetry indicators -} The parity eigenvalues $\xi_n(\Gamma_i) = \pm1$ of the occupied Bloch states at the four time-reversal invariant momenta (TRIMs) $\Gamma_{i=(n_1,n_2)} = \frac{1}{2}(n_1 \vv {b}_1 + n_2 \vv {b}_2)$, where $n_i = 0,1$ and $\mathbf {b}_i$ are primitive lattice vectors [See Fig.\,\ref{fig:sw_numbers}(a)]. They provide a symmetry indicator that diagnoses the 2DSWIs. The SW numbers $\nu_1$ and $\nu_2$ on the $k_1 = 0$ ($\pi$) and $k_2$ = 0 ($\pi$) loops satisfy
 \begin{align}
  (-1)^{\nu_1} = \xi_a \xi_d ~(\xi_b \xi_c); ~~(-1)^{\nu_2} = \xi_a \xi_b ~(\xi_d \xi_c),
 \label{eq:sw}
 \end{align}
where
\begin{align}
    \xi_a = \prod_n \xi_n(\Gamma_a).
\end{align}
($\nu_1\nu_2$) are well-defined if 
$\xi_a \xi_d = \xi_b \xi_c$ and $\xi_a \xi_b = \xi_d \xi_c$, guaranteed by the strong $\mathbb{Z}_2$ index $\nu_0= 0$ because
\begin{align}
    (-1)^{\nu_0} =  \xi_a\xi_b\xi_c\xi_d = 1.
\label{eq:parity}
\end{align}
Note that the strong index $\nu_0 = 0$ is also a necessary condition for an insulating phase 
since the strong index $\nu_0 = 1$ dictates the presence of the massless DPs in momentum space \cite{Kim15p036806}. Figure\,\ref{fig:sw_numbers}(b) shows possible configurations of parity eigenvalues compatible with $\nu_0 = 0$. We identify the 2DSWI as a $\mathcal{PT}$-symmetric topological insulator in vanishing SOC, characterized by $\nu_0 = 0$ and $(\nu_1\nu_2) \ne (00)$.

\cyan{2DSWI as a stack of SSH chains -} Similar to the weak topological insulators in three dimensions \cite{Moore07p121306, Fu07p045302}, which can be viewed as a stack of 2D quantum spin Hall insulators, the 2DSWI with $(\nu_1\nu_2) \ne (00)$ can be viewed as a stack of 1D SSH chains along $\vv G_\nu = \nu_1\vv b_1 +\nu_2\vv b_2$. In stark contrast to the 3D weak topological insulators, however, by stacking the SSH chains in two dimensions, every 1D line in the BZ has the first SW number. This results in the boundary mode, for example, at any edge momentum $\bar k_i$ when projected along the $k_j$-direction with $\nu_i = 1$, where $i$,$j$ = 1,2 and $i \ne j$. In general, the band topology of a 2DSWI manifests as 1D topological states at the $(\nu_1\nu_2)$ edge, as illustrated in Fig.\,\ref{fig:sw_numbers}(b). 

\begin{figure}
\includegraphics[width=0.5\textwidth]{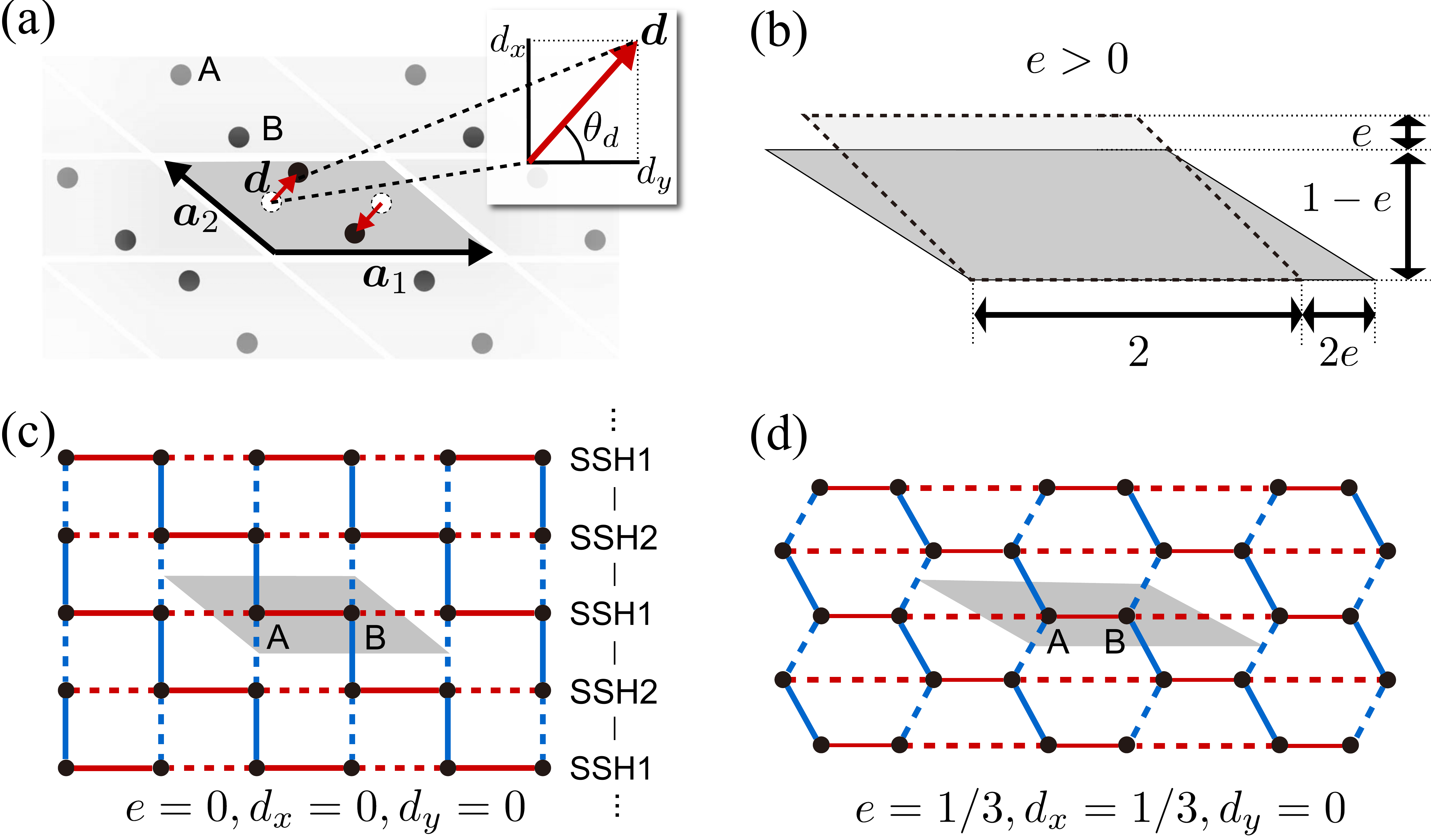}%
\caption{
Zhu-Prodan-Ahn model of the 2D coupled SSH chains. (a) A primitive unit cell of the 2D coupled SSH chains, indicated by a grey parallelogram. The staggered distortion, parameterized by $\vv d = (d_x,d_y)$, is applied to the $A$ and $B$ sublattices. (b) Unit cell under a strain $e \ne 0$.  Examples of the 2D coupled SSH chains: (c) Square lattice with $e = 0$ and $\vv d = 0$. (d) Honeycomb lattice with $e = 1/3$ and $\vv d = (1/3,0)$. The inter-chain and intra-chain dimerizations are delineated by red and blue-colored lines, respectively. A grey parallelogram indicates a primitive unit cell.
} 
\label{fig:ssh_chains}
\end{figure}

\cyan{Zhu-Prodan-Ahn model -} Using the ZPA model, we suggest a material realization of the 2DSWI in 2D coupled SSH chains. As shown in Fig.\,\ref{fig:ssh_chains}, a family of 2D coupled SSH chains are constructed by an alternating array of two SSH chains. The ZPA model considers inter- and intra-chain dimerizations, parameterized by $\vv d$ and $e$.  Here, $\vv d = (d_x,d_y)$ and $e$ describe an in-plane staggered distortion and a strain of the unit cell, respectively [See Figs.\,\ref{fig:ssh_chains}(a) and \ref{fig:ssh_chains}(b)]. This family of 2D coupled SSH chains includes various 2D lattices, such as a square lattice with $\vv d = 0$ and $e = 0$ [Figs.\,\ref{fig:ssh_chains}(c)] and a graphene-like honeycomb lattice with $e = 1/3$, $\vv d = \left(1/3, 0\right)$ [Fig.\,\ref{fig:ssh_chains}(d)].

A Hamiltonian, describing hopping of electrons in the 2D coupled SSH chains, is given by
\begin{align}  
\label{eq:hk}%
\mathcal{H} (\vv k) = h^\ast (\vv k) \sigma^+ +  h(\vv k) \sigma^-,
\end{align}
where $\sigma^\pm = \sigma_x \pm i \sigma_y$ are the Pauli matrices associated with the sublattices and $h(\vv k) = -t_1 - t_2 \mathrm e^{i k_2} - t_3 \mathrm e^{-i k_1} - t_4 \mathrm e^{-i(k_1 + k_2)}$. The hopping parameters $t_j$ ($j=1,2,3,4$) are given by $t_1 = 1 - e + 2d_x$, $t_2 = 1 + e + 2d_y$, $t_3 = 1 - e - 2d_x$, and $t_4 = 1 + e - 2d_y$, so that they describe the deformations. Diagonalizing Eq.\,(\ref{eq:hk}), one finds the following band dispersion:
\begin{align}  
\label{eq:energy}
E(\vv k)^2 & = 4 (1 + e^2 + 2d_x^2 + 2d_y^2) + 2 \left[ (1-e)^2 - 4 d_x^2 \right] \cos k_1  \nn
& + 4 \left[ (1-e^2) + 4 d_x d_y  \right] \cos k_2  \nn
& + 4 \left[ (1-e^2) - 4 d_x d_y  \right] \cos(k_1+k_2) \nn
& + 2 \left[ (1+e)^2 - 4d_y^2  \right] \cos(k_1+2k_2),
\end{align}
where $k_1 = \vv k \cdot \vv a_1 $ and $k_2 = \vv k \cdot \vv a_2 $ are dimensionless lattice momenta. 
In good agreement with the ZPA results \cite{Zhu19p041117}, the energy bands are calculated as fully gapped at $(e,d_x,d_y) = \left(1/3, 0, 1/3\right)$ and $(e,d_x,d_y) = \left(-1/3,1/3,0\right)$.

\begin{figure}			
\includegraphics[width=0.5\textwidth]{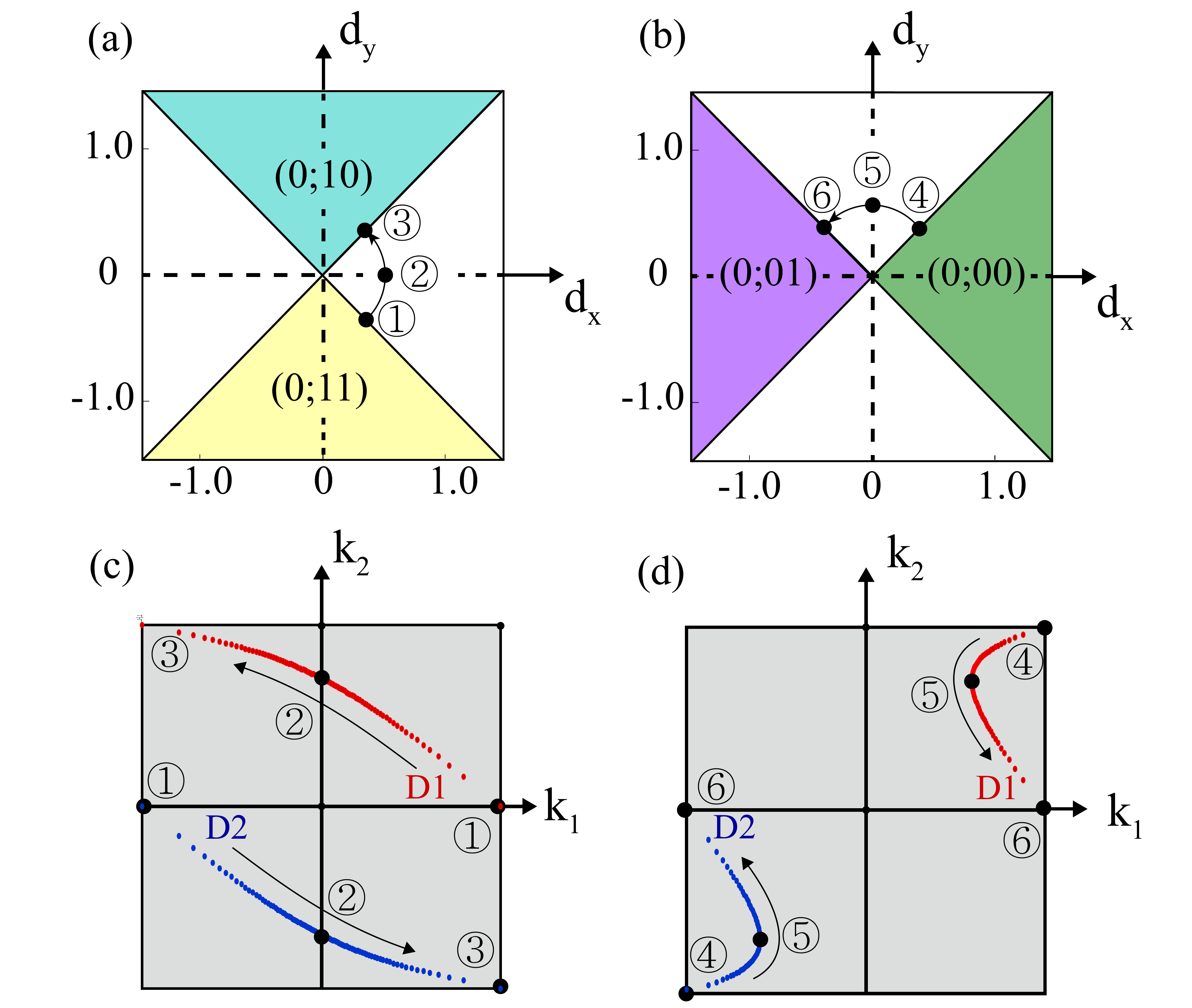}%
\caption{Topological phase diagrams with $\mathbb{Z}_2$ indices ($\nu_0;\nu_1\nu_2$) in $\vv d$-space. (a) e = 0.25. (b) e = -0.25. The colored (white) regions have the strong index $\nu_0 = 0$ ($\nu_0 = 1$). (c) Trajectories of the two DPs ($D1$ and $D2$) in momentum space along the path \textcircled{\small 1}, \textcircled{\small 2}, and \textcircled{\small 3} in (a). (d) Trajectories along the path \textcircled{\small 4}, \textcircled{\small 5}, and \textcircled{\small 6} in (b).
} 
\label{fig:phase_diagrams}
\end{figure}

\cyan{Topological phase diagrams -}  Topological phase diagrams in $\vv d$-space are calculated using the inversion-symmetry indicators ($\nu_0;\nu_1$$\nu_2$). Inversion operator, given by $\mathcal{P} = \sigma_x$, readily evaluates the strong index $\nu_0$ as
\begin{align}
    (-1)^{\nu_0} = \rm{sgn}\left[ e \sin\left(\theta_d - \frac{\pi}{4}\right)\sin\left(\theta_d+\frac{\pi}{4}\right) \right].
    \label{eq:nu_1}
\end{align}
where $\theta_d = \tan^{-1}(d_x/d_y)$. An insulating phase, facilitating the 2DSWIs, is allowed when $\nu_0 = 0$ or, equivalently, $e \sin\left(\theta_d - \frac{\pi}{4}\right)\sin\left(\theta_d+\frac{\pi}{4}\right) > 0$. A close inspection reveals that the bands are fully gapped when $\nu_0 = 0$. Using the symmetry indicators, we obtain the weak indices  ($\nu_1\nu_2$) as
\begin{align}
(-1)^{\nu_1} = - \rm{sgn}\left[ e\right]
\end{align}
and
\begin{align}
(-1)^{\nu_2} = \mathrm{sgn}\left[\sin\left(\theta_d+\frac{\pi}{4}\right) \right].
\end{align}
Figure\,\ref{fig:phase_diagrams} depicts the resultant phase diagrams. The phase boundaries always occur at $|d_y| = |d_x|$, irrespective of $|\vv d|$ and $e$. The sign of $e$ determines the metallic and insulating regions, depending on which there exist two kinds of topological phase diagrams. For $e > 0$, a metallic (insulating) phase occurs in $|\theta_d| < \pi/4$ and $|\theta_d - \pi| < \pi/4$ ($|\theta_d - \pi/2| < \pi/4$ and $|\theta_d + \pi/2| < \pi/4$) [Fig.\ref{fig:phase_diagrams}(a)]. For $e < 0$, the metallic and insulating phases switch the regions [Fig.\,\ref{fig:phase_diagrams}(b)]. The four insulating regions host distinct 2DSWIs with ($\nu_0$;$\nu_1$$\nu_2$) = (0;10), (0;11) [(0;01), and (0;00)] for $e > 0$ ($e < 0$), respectively.

\cyan{Topological phase transitions -}  The DPs that occur when $\nu_0 = 1$ are stable, existing in the finite regions of $\vv d$-space $e \sin\left(\theta_d - \frac{\pi}{4}\right)\sin\left(\theta_d+\frac{\pi}{4}\right) < 0$. The trajectories of the DPs as a function of the staggered dimerization angle $\theta_d$ capture the topological phase transition of the 2DSWIs, driven by the dimerization. Figures\,\ref{fig:phase_diagrams}(c) and \ref{fig:phase_diagrams}(d) show the trajectories of the DPs from $\theta_d = -\pi/4$ to $\pi/4$ for $e = 0.25$ and from $\theta_d = \pi/4$ to $3\pi/4$ for $e = -0.25$, respectively. When $e > 0$, the DPs traverse the BZ from $\Gamma_b$ to $\Gamma_c$, winding the 2D BZ twice (once) along the $k_1$- ($k_2$-)direction [Fig.\,\ref{fig:phase_diagrams}(c)]. On the other hand, when $e < 0$, the DPs travel from $\Gamma_a$ to $\Gamma_d$, winding the BZ once only along the $k_2$-direction [Fig.\,\ref{fig:phase_diagrams}(d)]. The BZ winding by DPs characterizes the 2DSWIs. The one-parameter families of real Hamiltonians on any $k_i$ line ($i$ = 1,2) defines a 1D topological insulator indexed by $\nu_i$. The weak $\mathbb{Z}_2$ indices ($\nu_1\nu_2$) can change by changing all the topological insulators at any $k_i \in [-\pi,\pi]$. This necessitates the existence of the Dirac nodal line that winds the BZ during the phase transition.

\begin{figure}
\includegraphics[width=0.5\textwidth]{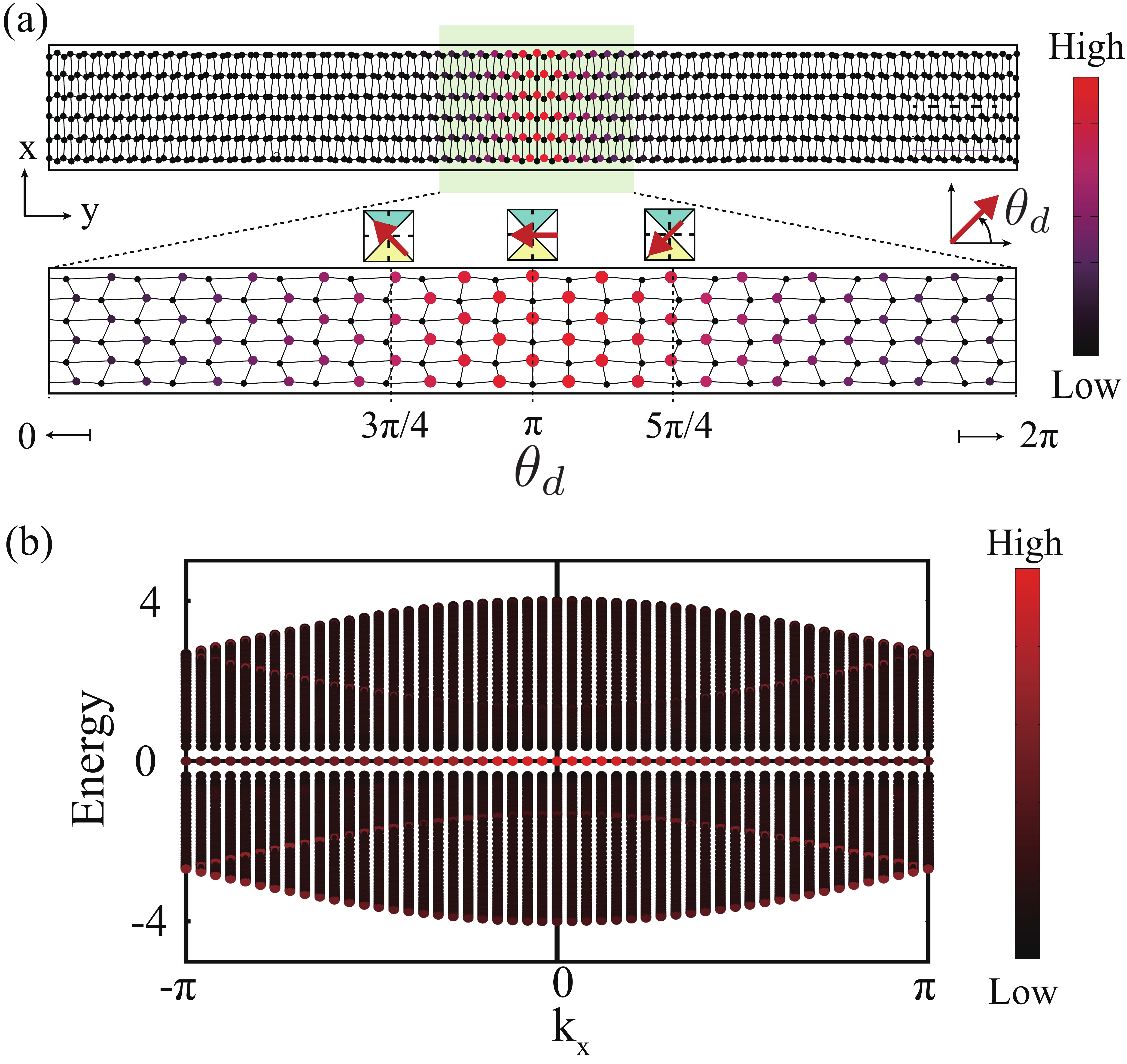}%
\caption{Atomic structure and electronic energy spectrum of the topological DW in 2D coupled SSH chains. (a) The atomic configuration of the DW geometry that maps $\theta_d \in [0,2\pi]$ to $y$ under the cell-periodicity along $x$ with $\vv d = e = 1/3$. The intensity represents the amplitude of the energy eigenstates with $E = 0$. $\theta_d$ is the staggered dimerization angle. (b) Energy spectrum of the DW geometry. The intensity represents the localization strength at the DW.}
\label{fig:dw_geo}
\end{figure}

\cyan{1D flat-band DW states -} We argue that the ZPA 1D boundary modes are a physical manifestation of the DPs that wind the BZ. ZPA have shown that the DWs of the 2D SSH chain host the one-dimensional flat bands \cite{Zhu19p041117}, which we reproduce in Fig.\,\ref{fig:dw_geo}. In view of the topological phase transition, the DW geometry [Fig.\,\ref{fig:dw_geo}(a)], in which the dimerization parameter $\vv d$ smoothly varies from $0$ to $2\pi$ along $y$, seamlessly visits the four electronic phases of the phase diagram [Fig.\,\ref{fig:phase_diagrams}(a)]. Therefore, the topological phase transition should be captured in between the insulating domains via the occurrence of two DPs that wind the edge BZ $\overline k_1 \in [-\pi,\pi]$ twice. Figure\,\ref{fig:dw_geo}(b) shows the electronic energy spectrum calculated from the DW structure. In good agreement with the previous results \cite{Zhu19p041117, Qian20p225501}, our calculations reproduce the flat bands at $E = 0$ in otherwise all gapped bulk states. We find that there exist two mid-gap states per DW [See Fig.\,\ref{fig:dw_geo}(b)], which cover the whole $\overline {k}_x$ edge BZ twice. Associated with the topological phase transition of $\nu_1$, these DW states can be considered as the projection of the DP trajectories from $\theta_d = -\pi/4$ to $\pi/4$ along $k_2$. Since the the DW states wind the $\overline {k}_1$ twice, the resulting change of $\nu_1$ should be zero, which is in line with the adjacent 2DSWI phases $(\nu_0;\nu_1 \nu_2) = (0;11)$ and $0;10)$. This supports that the 1D DW states are a physical manifestation of the DPs that wind the BZ. We note that the degeneracy of the zero modes are stabilized by the crystalline symmetry of the DW geometry as discussed by ZPA \cite{Zhu19p041117}.

It is interesting to note that the trajectories of the DPs and, thus, the 1D DW states can be viewed as a 1D projection of a Dirac nodal line that lives in higher three dimensions. A 3D BZ can be constructed by adding additional dimension from the polar coordinate $\theta_d \in [-\pi,\pi)$ to the 2D BZ. Viewed from the three dimensions ($k_1,k_2,\theta_d$), one can consider one-parameter families of the Hamiltonian $\mathcal{H}(k_1,k_2,\theta_d)$  in class AI \cite{Teo10p115120}, since satisfy $[\mathcal{H}(k_1,k_2,\theta_d),\mathcal{PT}] = 0$ and $\left(\mathcal{PT}\right)^{-1}\theta_d\mathcal{PT} = \theta_d$. Therefore, the strong index ($\nu_0 = 1$) is well-defined, dictating the presence of the Dirac nodal line that threads the 2D BZ at $\theta_d = 0$ ($\theta_d = \pi/2$) for $e > 0$ ($e < 0$). This idea may help access higher-dimensional band topology realized in a globally smooth spatially varying geometry of lower-dimensional materials.

\begin{figure}
\includegraphics[width=0.5\textwidth]{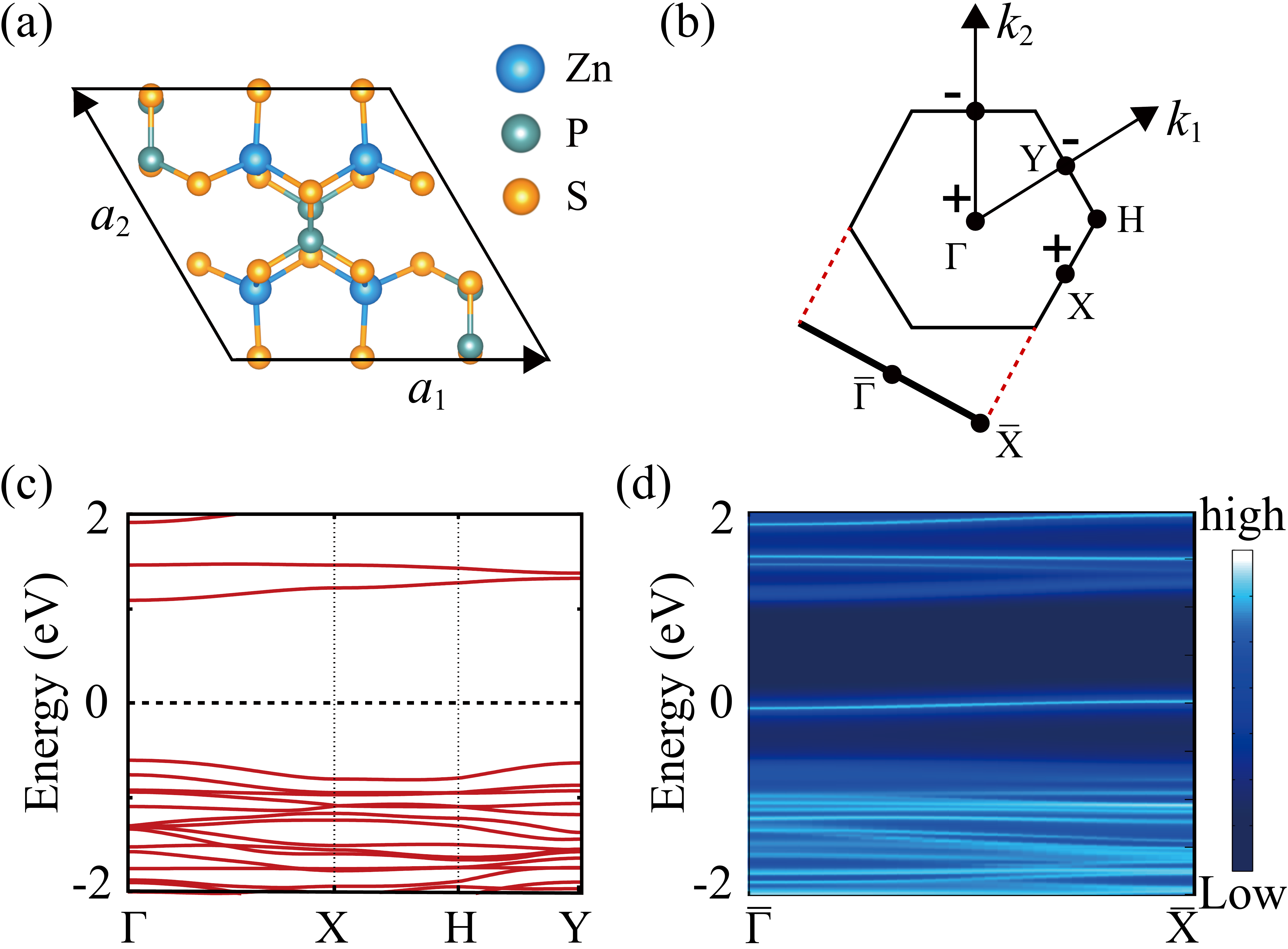}%
\caption{2DSWI phase in of Zn$_2$(PS$_3$)$_3$. (a) Atomic structure. A unit cell is indicated by solid box. (b) 2D bulk and 1D edge Brillouin zones. High-symmetric $\vv k$ points are indicated by filled circles. Parity eigenvalues are presented at the corresponding TRIMs, resulting in the nontrivial 2DSWI with the topological indices $(\nu_0;\nu_1\nu_2)=(0;11)$. (c) DFT energy bands without spin-orbit coupling. (d) Edge energy spectrum calculated along the projected BZ in (b). The higher the intensity, the stronger the concentration at the edges. Nearly flat edge states occurs as a consequence of the 2DSWI state in the bulk. 
}
\label{fig:ZPS}
\end{figure}
\cyan{Material realizations -} Finally, we use first-principles calculations based on density functional theory (DFT) to predict the materials that realize the first 2DSWI phase \footnote{See the Supplemental Material for the detailed calculations methods and the list of the materials that realize the first 2DSWI}. Using the suggested indices indicated by parity eigenvalues, eleven known materials are identified as the first 2DSWI \footnote{See the Supplemental Material for the list of the materials.}. As a representative example, here we demonstrate the 2DSWI in Zn$_2$(PS$_3$)$_3$. Figure\,\ref{fig:ZPS}(a) shows the monoclinic atomic structure of Zn$_2$(PS$_3$)$_3$ in space group C2/m (\# 12), which includes most importantly inversion symmetry. Zn$_2$(PS$_3$)$_3$ has a well-defined band gap throughout the entire BZ as shown in Fig.\,\ref{fig:ZPS}. Evaluating the parity eigenvalues, we find the weak indices $(\nu_1\nu_2) = (11)$ for the delineated atomic structure. The nontrivial indices dictate the presence of the edge states along the $\vv k_1 + \vv k_2$ direction, confirmed by our DFT calculations in Fig.\,\ref{fig:ZPS}(c). Our DFT first 2DSWI belongs to an important class of materials, realized in realistic materials with intriguing physical manifestation as nearly-flat edge states. Furthermore, our first-principles calculations show that the parity eigenvalues that we suggest in the present study serve as a symmetry indicator to discern the 2DSWI, potentially leading to further discovery of material realizations and the experimental observations.

\cyan{Conclusion -} In summary, we have demonstrated that the combination of inversion and time-reversal symmetries allows for the $\mathbb{Z}_2$ classification of topological insulators under vanishing spin-orbit interactions. The proposed topological insulators are characterized by the strong $\mathbb{Z}_2$ index $\nu_0 = 0$ and the weak $\mathbb{Z}_2$ indices $(\nu_1\nu_2) \ne (00)$, indicated by parity eigenvalues at four TRIMs. Diagnosing with the topological indices, we have shed light on the centrosymmetric 2D coupled SSH chains based on the ZPA model as a material realization of the proposed topological insulators, dubbed the two-dimensional Stiefel-Whitney insulators (2DSWIs). The 2DSWI features 1D topological boundary modes as a physical manifestation of their characteristic topological phase transitions. The proposed inversion-symmetry indicators can help identify the 2DSWIs in solid-state materials. Hopefully, our results simulate further experimental and theoretical studies and lead to the discover a solid-state platform that realizes topological flat bands.

\begin{acknowledgments}
This work was supported by National Research Foundation (NRF) of Korea under the program of Basic Research Laboratory (2020R1A4A3079707). Y.K. acknowledges the support from the NRF Grants (2021R1A2C1013871). The computational resource was provided from the Korea Institute of Science and Technology Information (KISTI) (KSC-2021-CRE-0116).
\end{acknowledgments}

\bibliography{refs}

\begin{thebibliography}{60}%
\makeatletter
\providecommand \@ifxundefined [1]{%
 \@ifx{#1\undefined}
}%
\providecommand \@ifnum [1]{%
 \ifnum #1\expandafter \@firstoftwo
 \else \expandafter \@secondoftwo
 \fi
}%
\providecommand \@ifx [1]{%
 \ifx #1\expandafter \@firstoftwo
 \else \expandafter \@secondoftwo
 \fi
}%
\providecommand \natexlab [1]{#1}%
\providecommand \enquote  [1]{``#1''}%
\providecommand \bibnamefont  [1]{#1}%
\providecommand \bibfnamefont [1]{#1}%
\providecommand \citenamefont [1]{#1}%
\providecommand \href@noop [0]{\@secondoftwo}%
\providecommand \href [0]{\begingroup \@sanitize@url \@href}%
\providecommand \@href[1]{\@@startlink{#1}\@@href}%
\providecommand \@@href[1]{\endgroup#1\@@endlink}%
\providecommand \@sanitize@url [0]{\catcode `\\12\catcode `\$12\catcode
  `\&12\catcode `\#12\catcode `\^12\catcode `\_12\catcode `\%12\relax}%
\providecommand \@@startlink[1]{}%
\providecommand \@@endlink[0]{}%
\providecommand \url  [0]{\begingroup\@sanitize@url \@url }%
\providecommand \@url [1]{\endgroup\@href {#1}{\urlprefix }}%
\providecommand \urlprefix  [0]{URL }%
\providecommand \Eprint [0]{\href }%
\providecommand \doibase [0]{http://dx.doi.org/}%
\providecommand \selectlanguage [0]{\@gobble}%
\providecommand \bibinfo  [0]{\@secondoftwo}%
\providecommand \bibfield  [0]{\@secondoftwo}%
\providecommand \translation [1]{[#1]}%
\providecommand \BibitemOpen [0]{}%
\providecommand \bibitemStop [0]{}%
\providecommand \bibitemNoStop [0]{.\EOS\space}%
\providecommand \EOS [0]{\spacefactor3000\relax}%
\providecommand \BibitemShut  [1]{\csname bibitem#1\endcsname}%
\let\auto@bib@innerbib\@empty
\bibitem [{\citenamefont {Wu}\ \emph {et~al.}(2007)\citenamefont {Wu},
  \citenamefont {Bergman}, \citenamefont {Balents},\ and\ \citenamefont
  {Das~Sarma}}]{Wu07p070401}%
  \BibitemOpen
  \bibfield  {author} {\bibinfo {author} {\bibfnamefont {C.}~\bibnamefont
  {Wu}}, \bibinfo {author} {\bibfnamefont {D.}~\bibnamefont {Bergman}},
  \bibinfo {author} {\bibfnamefont {L.}~\bibnamefont {Balents}}, \ and\
  \bibinfo {author} {\bibfnamefont {S.}~\bibnamefont {Das~Sarma}},\ }\href
  {\doibase 10.1103/PhysRevLett.99.070401} {\bibfield  {journal} {\bibinfo
  {journal} {Phys. Rev. Lett.}\ }\textbf {\bibinfo {volume} {99}},\ \bibinfo
  {pages} {070401} (\bibinfo {year} {2007})}\BibitemShut {NoStop}%
\bibitem [{\citenamefont {Wong}\ \emph {et~al.}(2013)\citenamefont {Wong},
  \citenamefont {Liu}, \citenamefont {Law},\ and\ \citenamefont
  {Lee}}]{Wong13p060504}%
  \BibitemOpen
  \bibfield  {author} {\bibinfo {author} {\bibfnamefont {C.~L.~M.}\
  \bibnamefont {Wong}}, \bibinfo {author} {\bibfnamefont {J.}~\bibnamefont
  {Liu}}, \bibinfo {author} {\bibfnamefont {K.~T.}\ \bibnamefont {Law}}, \ and\
  \bibinfo {author} {\bibfnamefont {P.~A.}\ \bibnamefont {Lee}},\ }\href
  {\doibase 10.1103/PhysRevB.88.060504} {\bibfield  {journal} {\bibinfo
  {journal} {Phys. Rev. B}\ }\textbf {\bibinfo {volume} {88}},\ \bibinfo
  {pages} {060504} (\bibinfo {year} {2013})}\BibitemShut {NoStop}%
\bibitem [{\citenamefont {Biesenthal}\ \emph {et~al.}(2019)\citenamefont
  {Biesenthal}, \citenamefont {Kremer}, \citenamefont {Heinrich},\ and\
  \citenamefont {Szameit}}]{Biesenthal19p183601}%
  \BibitemOpen
  \bibfield  {author} {\bibinfo {author} {\bibfnamefont {T.}~\bibnamefont
  {Biesenthal}}, \bibinfo {author} {\bibfnamefont {M.}~\bibnamefont {Kremer}},
  \bibinfo {author} {\bibfnamefont {M.}~\bibnamefont {Heinrich}}, \ and\
  \bibinfo {author} {\bibfnamefont {A.}~\bibnamefont {Szameit}},\ }\href
  {\doibase 10.1103/PhysRevLett.123.183601} {\bibfield  {journal} {\bibinfo
  {journal} {Phys. Rev. Lett.}\ }\textbf {\bibinfo {volume} {123}},\ \bibinfo
  {pages} {183601} (\bibinfo {year} {2019})}\BibitemShut {NoStop}%
\bibitem [{\citenamefont {Balents}\ \emph {et~al.}(2020)\citenamefont
  {Balents}, \citenamefont {Dean}, \citenamefont {Efetov},\ and\ \citenamefont
  {Young}}]{Balents20p725}%
  \BibitemOpen
  \bibfield  {author} {\bibinfo {author} {\bibfnamefont {L.}~\bibnamefont
  {Balents}}, \bibinfo {author} {\bibfnamefont {C.~R.}\ \bibnamefont {Dean}},
  \bibinfo {author} {\bibfnamefont {D.~K.}\ \bibnamefont {Efetov}}, \ and\
  \bibinfo {author} {\bibfnamefont {A.~F.}\ \bibnamefont {Young}},\ }\href
  {\doibase 10.1038/s41567-020-0906-9} {\bibfield  {journal} {\bibinfo
  {journal} {Nature Physics}\ }\textbf {\bibinfo {volume} {16}},\ \bibinfo
  {pages} {725} (\bibinfo {year} {2020})}\BibitemShut {NoStop}%
\bibitem [{\citenamefont {Lopes~dos Santos}\ \emph {et~al.}(2007)\citenamefont
  {Lopes~dos Santos}, \citenamefont {Peres},\ and\ \citenamefont
  {Castro~Neto}}]{Santos07p256802}%
  \BibitemOpen
  \bibfield  {author} {\bibinfo {author} {\bibfnamefont {J.~M.~B.}\
  \bibnamefont {Lopes~dos Santos}}, \bibinfo {author} {\bibfnamefont
  {N.~M.~R.}\ \bibnamefont {Peres}}, \ and\ \bibinfo {author} {\bibfnamefont
  {A.~H.}\ \bibnamefont {Castro~Neto}},\ }\href {\doibase
  10.1103/PhysRevLett.99.256802} {\bibfield  {journal} {\bibinfo  {journal}
  {Phys. Rev. Lett.}\ }\textbf {\bibinfo {volume} {99}},\ \bibinfo {pages}
  {256802} (\bibinfo {year} {2007})}\BibitemShut {NoStop}%
\bibitem [{\citenamefont {Dean}\ \emph {et~al.}(2010)\citenamefont {Dean},
  \citenamefont {Young}, \citenamefont {Meric}, \citenamefont {Lee},
  \citenamefont {Wang}, \citenamefont {Sorgenfrei}, \citenamefont {Watanabe},
  \citenamefont {Taniguchi}, \citenamefont {Kim}, \citenamefont {Shepard},\
  and\ \citenamefont {Hone}}]{Dean10p722}%
  \BibitemOpen
  \bibfield  {author} {\bibinfo {author} {\bibfnamefont {C.~R.}\ \bibnamefont
  {Dean}}, \bibinfo {author} {\bibfnamefont {A.~F.}\ \bibnamefont {Young}},
  \bibinfo {author} {\bibfnamefont {I.}~\bibnamefont {Meric}}, \bibinfo
  {author} {\bibfnamefont {C.}~\bibnamefont {Lee}}, \bibinfo {author}
  {\bibfnamefont {L.}~\bibnamefont {Wang}}, \bibinfo {author} {\bibfnamefont
  {S.}~\bibnamefont {Sorgenfrei}}, \bibinfo {author} {\bibfnamefont
  {K.}~\bibnamefont {Watanabe}}, \bibinfo {author} {\bibfnamefont
  {T.}~\bibnamefont {Taniguchi}}, \bibinfo {author} {\bibfnamefont
  {P.}~\bibnamefont {Kim}}, \bibinfo {author} {\bibfnamefont {K.~L.}\
  \bibnamefont {Shepard}}, \ and\ \bibinfo {author} {\bibfnamefont
  {J.}~\bibnamefont {Hone}},\ }\href {\doibase 10.1038/nnano.2010.172}
  {\bibfield  {journal} {\bibinfo  {journal} {Nature Nanotechnology}\ }\textbf
  {\bibinfo {volume} {5}},\ \bibinfo {pages} {722} (\bibinfo {year}
  {2010})}\BibitemShut {NoStop}%
\bibitem [{\citenamefont {Su\'arez~Morell}\ \emph {et~al.}(2010)\citenamefont
  {Su\'arez~Morell}, \citenamefont {Correa}, \citenamefont {Vargas},
  \citenamefont {Pacheco},\ and\ \citenamefont {Barticevic}}]{Morell10p121407}%
  \BibitemOpen
  \bibfield  {author} {\bibinfo {author} {\bibfnamefont {E.}~\bibnamefont
  {Su\'arez~Morell}}, \bibinfo {author} {\bibfnamefont {J.~D.}\ \bibnamefont
  {Correa}}, \bibinfo {author} {\bibfnamefont {P.}~\bibnamefont {Vargas}},
  \bibinfo {author} {\bibfnamefont {M.}~\bibnamefont {Pacheco}}, \ and\
  \bibinfo {author} {\bibfnamefont {Z.}~\bibnamefont {Barticevic}},\ }\href
  {\doibase 10.1103/PhysRevB.82.121407} {\bibfield  {journal} {\bibinfo
  {journal} {Phys. Rev. B}\ }\textbf {\bibinfo {volume} {82}},\ \bibinfo
  {pages} {121407(R)} (\bibinfo {year} {2010})}\BibitemShut {NoStop}%
\bibitem [{\citenamefont {Bistritzer}\ and\ \citenamefont
  {MacDonald}(2011)}]{Bistritzer11p12233}%
  \BibitemOpen
  \bibfield  {author} {\bibinfo {author} {\bibfnamefont {R.}~\bibnamefont
  {Bistritzer}}\ and\ \bibinfo {author} {\bibfnamefont {A.~H.}\ \bibnamefont
  {MacDonald}},\ }\href {\doibase 10.1073/pnas.1108174108} {\bibfield
  {journal} {\bibinfo  {journal} {Proceedings of the National Academy of
  Sciences}\ }\textbf {\bibinfo {volume} {108}},\ \bibinfo {pages} {12233}
  (\bibinfo {year} {2011})}\BibitemShut {NoStop}%
\bibitem [{\citenamefont {Wang}\ \emph {et~al.}(2013)\citenamefont {Wang},
  \citenamefont {Meric}, \citenamefont {Huang}, \citenamefont {Gao},
  \citenamefont {Gao}, \citenamefont {Tran}, \citenamefont {Taniguchi},
  \citenamefont {Watanabe}, \citenamefont {Campos}, \citenamefont {Muller},
  \citenamefont {Guo}, \citenamefont {Kim}, \citenamefont {Hone}, \citenamefont
  {Shepard},\ and\ \citenamefont {Dean}}]{Wang13p614}%
  \BibitemOpen
  \bibfield  {author} {\bibinfo {author} {\bibfnamefont {L.}~\bibnamefont
  {Wang}}, \bibinfo {author} {\bibfnamefont {I.}~\bibnamefont {Meric}},
  \bibinfo {author} {\bibfnamefont {P.~Y.}\ \bibnamefont {Huang}}, \bibinfo
  {author} {\bibfnamefont {Q.}~\bibnamefont {Gao}}, \bibinfo {author}
  {\bibfnamefont {Y.}~\bibnamefont {Gao}}, \bibinfo {author} {\bibfnamefont
  {H.}~\bibnamefont {Tran}}, \bibinfo {author} {\bibfnamefont {T.}~\bibnamefont
  {Taniguchi}}, \bibinfo {author} {\bibfnamefont {K.}~\bibnamefont {Watanabe}},
  \bibinfo {author} {\bibfnamefont {L.~M.}\ \bibnamefont {Campos}}, \bibinfo
  {author} {\bibfnamefont {D.~A.}\ \bibnamefont {Muller}}, \bibinfo {author}
  {\bibfnamefont {J.}~\bibnamefont {Guo}}, \bibinfo {author} {\bibfnamefont
  {P.}~\bibnamefont {Kim}}, \bibinfo {author} {\bibfnamefont {J.}~\bibnamefont
  {Hone}}, \bibinfo {author} {\bibfnamefont {K.~L.}\ \bibnamefont {Shepard}}, \
  and\ \bibinfo {author} {\bibfnamefont {C.~R.}\ \bibnamefont {Dean}},\ }\href
  {\doibase 10.1126/science.1244358} {\bibfield  {journal} {\bibinfo  {journal}
  {Science}\ }\textbf {\bibinfo {volume} {342}},\ \bibinfo {pages} {614}
  (\bibinfo {year} {2013})}\BibitemShut {NoStop}%
\bibitem [{\citenamefont {Kim}\ \emph {et~al.}(2016)\citenamefont {Kim},
  \citenamefont {Yankowitz}, \citenamefont {Fallahazad}, \citenamefont {Kang},
  \citenamefont {Movva}, \citenamefont {Huang}, \citenamefont {Larentis},
  \citenamefont {Corbet}, \citenamefont {Taniguchi}, \citenamefont {Watanabe},
  \citenamefont {Banerjee}, \citenamefont {LeRoy},\ and\ \citenamefont
  {Tutuc}}]{Kim16p1989}%
  \BibitemOpen
  \bibfield  {author} {\bibinfo {author} {\bibfnamefont {K.}~\bibnamefont
  {Kim}}, \bibinfo {author} {\bibfnamefont {M.}~\bibnamefont {Yankowitz}},
  \bibinfo {author} {\bibfnamefont {B.}~\bibnamefont {Fallahazad}}, \bibinfo
  {author} {\bibfnamefont {S.}~\bibnamefont {Kang}}, \bibinfo {author}
  {\bibfnamefont {H.~C.~P.}\ \bibnamefont {Movva}}, \bibinfo {author}
  {\bibfnamefont {S.}~\bibnamefont {Huang}}, \bibinfo {author} {\bibfnamefont
  {S.}~\bibnamefont {Larentis}}, \bibinfo {author} {\bibfnamefont {C.~M.}\
  \bibnamefont {Corbet}}, \bibinfo {author} {\bibfnamefont {T.}~\bibnamefont
  {Taniguchi}}, \bibinfo {author} {\bibfnamefont {K.}~\bibnamefont {Watanabe}},
  \bibinfo {author} {\bibfnamefont {S.~K.}\ \bibnamefont {Banerjee}}, \bibinfo
  {author} {\bibfnamefont {B.~J.}\ \bibnamefont {LeRoy}}, \ and\ \bibinfo
  {author} {\bibfnamefont {E.}~\bibnamefont {Tutuc}},\ }\href {\doibase
  10.1021/acs.nanolett.5b05263} {\bibfield  {journal} {\bibinfo  {journal}
  {Nano Letters}\ }\textbf {\bibinfo {volume} {16}},\ \bibinfo {pages} {1989}
  (\bibinfo {year} {2016})}\BibitemShut {NoStop}%
\bibitem [{\citenamefont {Cao}\ \emph {et~al.}(2018{\natexlab{a}})\citenamefont
  {Cao}, \citenamefont {Fatemi}, \citenamefont {Fang}, \citenamefont
  {Watanabe}, \citenamefont {Taniguchi}, \citenamefont {Kaxiras},\ and\
  \citenamefont {Jarillo-Herrero}}]{Cao18p43}%
  \BibitemOpen
  \bibfield  {author} {\bibinfo {author} {\bibfnamefont {Y.}~\bibnamefont
  {Cao}}, \bibinfo {author} {\bibfnamefont {V.}~\bibnamefont {Fatemi}},
  \bibinfo {author} {\bibfnamefont {S.}~\bibnamefont {Fang}}, \bibinfo {author}
  {\bibfnamefont {K.}~\bibnamefont {Watanabe}}, \bibinfo {author}
  {\bibfnamefont {T.}~\bibnamefont {Taniguchi}}, \bibinfo {author}
  {\bibfnamefont {E.}~\bibnamefont {Kaxiras}}, \ and\ \bibinfo {author}
  {\bibfnamefont {P.}~\bibnamefont {Jarillo-Herrero}},\ }\href {\doibase
  10.1038/nature26160} {\bibfield  {journal} {\bibinfo  {journal} {Nature}\
  }\textbf {\bibinfo {volume} {556}},\ \bibinfo {pages} {43} (\bibinfo {year}
  {2018}{\natexlab{a}})}\BibitemShut {NoStop}%
\bibitem [{\citenamefont {Cao}\ \emph {et~al.}(2018{\natexlab{b}})\citenamefont
  {Cao}, \citenamefont {Fatemi}, \citenamefont {Demir}, \citenamefont {Fang},
  \citenamefont {Tomarken}, \citenamefont {Luo}, \citenamefont
  {Sanchez-Yamagishi}, \citenamefont {Watanabe}, \citenamefont {Taniguchi},
  \citenamefont {Kaxiras}, \citenamefont {Ashoori},\ and\ \citenamefont
  {Jarillo-Herrero}}]{Cao18p80}%
  \BibitemOpen
  \bibfield  {author} {\bibinfo {author} {\bibfnamefont {Y.}~\bibnamefont
  {Cao}}, \bibinfo {author} {\bibfnamefont {V.}~\bibnamefont {Fatemi}},
  \bibinfo {author} {\bibfnamefont {A.}~\bibnamefont {Demir}}, \bibinfo
  {author} {\bibfnamefont {S.}~\bibnamefont {Fang}}, \bibinfo {author}
  {\bibfnamefont {S.~L.}\ \bibnamefont {Tomarken}}, \bibinfo {author}
  {\bibfnamefont {J.~Y.}\ \bibnamefont {Luo}}, \bibinfo {author} {\bibfnamefont
  {J.~D.}\ \bibnamefont {Sanchez-Yamagishi}}, \bibinfo {author} {\bibfnamefont
  {K.}~\bibnamefont {Watanabe}}, \bibinfo {author} {\bibfnamefont
  {T.}~\bibnamefont {Taniguchi}}, \bibinfo {author} {\bibfnamefont
  {E.}~\bibnamefont {Kaxiras}}, \bibinfo {author} {\bibfnamefont {R.~C.}\
  \bibnamefont {Ashoori}}, \ and\ \bibinfo {author} {\bibfnamefont
  {P.}~\bibnamefont {Jarillo-Herrero}},\ }\href {\doibase 10.1038/nature26154}
  {\bibfield  {journal} {\bibinfo  {journal} {Nature}\ }\textbf {\bibinfo
  {volume} {556}},\ \bibinfo {pages} {80} (\bibinfo {year}
  {2018}{\natexlab{b}})}\BibitemShut {NoStop}%
\bibitem [{\citenamefont {Song}\ \emph {et~al.}(2019)\citenamefont {Song},
  \citenamefont {Wang}, \citenamefont {Shi}, \citenamefont {Li}, \citenamefont
  {Fang},\ and\ \citenamefont {Bernevig}}]{Song19p036401}%
  \BibitemOpen
  \bibfield  {author} {\bibinfo {author} {\bibfnamefont {Z.}~\bibnamefont
  {Song}}, \bibinfo {author} {\bibfnamefont {Z.}~\bibnamefont {Wang}}, \bibinfo
  {author} {\bibfnamefont {W.}~\bibnamefont {Shi}}, \bibinfo {author}
  {\bibfnamefont {G.}~\bibnamefont {Li}}, \bibinfo {author} {\bibfnamefont
  {C.}~\bibnamefont {Fang}}, \ and\ \bibinfo {author} {\bibfnamefont {B.~A.}\
  \bibnamefont {Bernevig}},\ }\href {\doibase 10.1103/PhysRevLett.123.036401}
  {\bibfield  {journal} {\bibinfo  {journal} {Phys. Rev. Lett.}\ }\textbf
  {\bibinfo {volume} {123}},\ \bibinfo {pages} {036401} (\bibinfo {year}
  {2019})}\BibitemShut {NoStop}%
\bibitem [{\citenamefont {Brihuega}\ \emph {et~al.}(2012)\citenamefont
  {Brihuega}, \citenamefont {Mallet}, \citenamefont {Gonz\'alez-Herrero},
  \citenamefont {Trambly~de Laissardi\`ere}, \citenamefont {Ugeda},
  \citenamefont {Magaud}, \citenamefont {G\'omez-Rodr\'{\i}guez}, \citenamefont
  {Yndur\'ain},\ and\ \citenamefont {Veuillen}}]{Brihuega19p196802}%
  \BibitemOpen
  \bibfield  {author} {\bibinfo {author} {\bibfnamefont {I.}~\bibnamefont
  {Brihuega}}, \bibinfo {author} {\bibfnamefont {P.}~\bibnamefont {Mallet}},
  \bibinfo {author} {\bibfnamefont {H.}~\bibnamefont {Gonz\'alez-Herrero}},
  \bibinfo {author} {\bibfnamefont {G.}~\bibnamefont {Trambly~de
  Laissardi\`ere}}, \bibinfo {author} {\bibfnamefont {M.~M.}\ \bibnamefont
  {Ugeda}}, \bibinfo {author} {\bibfnamefont {L.}~\bibnamefont {Magaud}},
  \bibinfo {author} {\bibfnamefont {J.~M.}\ \bibnamefont
  {G\'omez-Rodr\'{\i}guez}}, \bibinfo {author} {\bibfnamefont {F.}~\bibnamefont
  {Yndur\'ain}}, \ and\ \bibinfo {author} {\bibfnamefont {J.-Y.}\ \bibnamefont
  {Veuillen}},\ }\href {\doibase 10.1103/PhysRevLett.109.196802} {\bibfield
  {journal} {\bibinfo  {journal} {Phys. Rev. Lett.}\ }\textbf {\bibinfo
  {volume} {109}},\ \bibinfo {pages} {196802} (\bibinfo {year}
  {2012})}\BibitemShut {NoStop}%
\bibitem [{\citenamefont {Heikkil\"{a}}\ \emph {et~al.}(2011)\citenamefont
  {Heikkil\"{a}}, \citenamefont {Kopnin},\ and\ \citenamefont
  {Volovik}}]{Heikkil011p233}%
  \BibitemOpen
  \bibfield  {author} {\bibinfo {author} {\bibfnamefont {T.~T.}\ \bibnamefont
  {Heikkil\"{a}}}, \bibinfo {author} {\bibfnamefont {N.~B.}\ \bibnamefont
  {Kopnin}}, \ and\ \bibinfo {author} {\bibfnamefont {G.~E.}\ \bibnamefont
  {Volovik}},\ }\href {\doibase 10.1134/s0021364011150045} {\bibfield
  {journal} {\bibinfo  {journal} {{JETP} Letters}\ }\textbf {\bibinfo {volume}
  {94}},\ \bibinfo {pages} {233} (\bibinfo {year} {2011})}\BibitemShut
  {NoStop}%
\bibitem [{\citenamefont {Bi}\ \emph {et~al.}(2019)\citenamefont {Bi},
  \citenamefont {Yuan},\ and\ \citenamefont {Fu}}]{Fu19p035448}%
  \BibitemOpen
  \bibfield  {author} {\bibinfo {author} {\bibfnamefont {Z.}~\bibnamefont
  {Bi}}, \bibinfo {author} {\bibfnamefont {N.~F.~Q.}\ \bibnamefont {Yuan}}, \
  and\ \bibinfo {author} {\bibfnamefont {L.}~\bibnamefont {Fu}},\ }\href
  {\doibase 10.1103/PhysRevB.100.035448} {\bibfield  {journal} {\bibinfo
  {journal} {Phys. Rev. B}\ }\textbf {\bibinfo {volume} {100}},\ \bibinfo
  {pages} {035448} (\bibinfo {year} {2019})}\BibitemShut {NoStop}%
\bibitem [{\citenamefont {Hwang}\ \emph {et~al.}(2021)\citenamefont {Hwang},
  \citenamefont {Rhim},\ and\ \citenamefont {Yang}}]{hwang21p2105.14919}%
  \BibitemOpen
  \bibfield  {author} {\bibinfo {author} {\bibfnamefont {Y.}~\bibnamefont
  {Hwang}}, \bibinfo {author} {\bibfnamefont {J.-W.}\ \bibnamefont {Rhim}}, \
  and\ \bibinfo {author} {\bibfnamefont {B.-J.}\ \bibnamefont {Yang}},\
  }\href@noop {} {\enquote {\bibinfo {title} {Flat bands with band crossings
  enforced by symmetry representation},}\ } (\bibinfo {year} {2021}),\ \Eprint
  {http://arxiv.org/abs/2105.14919} {arXiv:2105.14919 [cond-mat.mes-hall]}
  \BibitemShut {NoStop}%
\bibitem [{\citenamefont {Zhu}\ \emph {et~al.}(2019)\citenamefont {Zhu},
  \citenamefont {Prodan},\ and\ \citenamefont {Ahn}}]{Zhu19p041117}%
  \BibitemOpen
  \bibfield  {author} {\bibinfo {author} {\bibfnamefont {L.}~\bibnamefont
  {Zhu}}, \bibinfo {author} {\bibfnamefont {E.}~\bibnamefont {Prodan}}, \ and\
  \bibinfo {author} {\bibfnamefont {K.~H.}\ \bibnamefont {Ahn}},\ }\href
  {\doibase 10.1103/PhysRevB.99.041117} {\bibfield  {journal} {\bibinfo
  {journal} {Phys. Rev. B}\ }\textbf {\bibinfo {volume} {99}},\ \bibinfo
  {pages} {041117} (\bibinfo {year} {2019})}\BibitemShut {NoStop}%
\bibitem [{\citenamefont {Su}\ \emph {et~al.}(1979)\citenamefont {Su},
  \citenamefont {Schrieffer},\ and\ \citenamefont {Heeger}}]{SSH79p1698}%
  \BibitemOpen
  \bibfield  {author} {\bibinfo {author} {\bibfnamefont {W.~P.}\ \bibnamefont
  {Su}}, \bibinfo {author} {\bibfnamefont {J.~R.}\ \bibnamefont {Schrieffer}},
  \ and\ \bibinfo {author} {\bibfnamefont {A.~J.}\ \bibnamefont {Heeger}},\
  }\href {\doibase 10.1103/PhysRevLett.42.1698} {\bibfield  {journal} {\bibinfo
   {journal} {Phys. Rev. Lett.}\ }\textbf {\bibinfo {volume} {42}},\ \bibinfo
  {pages} {1698} (\bibinfo {year} {1979})}\BibitemShut {NoStop}%
\bibitem [{\citenamefont {Su}\ \emph {et~al.}(1980)\citenamefont {Su},
  \citenamefont {Schrieffer},\ and\ \citenamefont {Heeger}}]{PhysRevB.22.2099}%
  \BibitemOpen
  \bibfield  {author} {\bibinfo {author} {\bibfnamefont {W.~P.}\ \bibnamefont
  {Su}}, \bibinfo {author} {\bibfnamefont {J.~R.}\ \bibnamefont {Schrieffer}},
  \ and\ \bibinfo {author} {\bibfnamefont {A.~J.}\ \bibnamefont {Heeger}},\
  }\href {\doibase 10.1103/PhysRevB.22.2099} {\bibfield  {journal} {\bibinfo
  {journal} {Phys. Rev. B}\ }\textbf {\bibinfo {volume} {22}},\ \bibinfo
  {pages} {2099} (\bibinfo {year} {1980})}\BibitemShut {NoStop}%
\bibitem [{\citenamefont {Heeger}\ \emph {et~al.}(1988)\citenamefont {Heeger},
  \citenamefont {Kivelson}, \citenamefont {Schrieffer},\ and\ \citenamefont
  {Su}}]{RevModPhys.60.781}%
  \BibitemOpen
  \bibfield  {author} {\bibinfo {author} {\bibfnamefont {A.~J.}\ \bibnamefont
  {Heeger}}, \bibinfo {author} {\bibfnamefont {S.}~\bibnamefont {Kivelson}},
  \bibinfo {author} {\bibfnamefont {J.~R.}\ \bibnamefont {Schrieffer}}, \ and\
  \bibinfo {author} {\bibfnamefont {W.~P.}\ \bibnamefont {Su}},\ }\href
  {\doibase 10.1103/RevModPhys.60.781} {\bibfield  {journal} {\bibinfo
  {journal} {Rev. Mod. Phys.}\ }\textbf {\bibinfo {volume} {60}},\ \bibinfo
  {pages} {781} (\bibinfo {year} {1988})}\BibitemShut {NoStop}%
\bibitem [{\citenamefont {Qian}\ \emph {et~al.}(2020)\citenamefont {Qian},
  \citenamefont {Zhu}, \citenamefont {Ahn},\ and\ \citenamefont
  {Prodan}}]{Qian20p225501}%
  \BibitemOpen
  \bibfield  {author} {\bibinfo {author} {\bibfnamefont {K.}~\bibnamefont
  {Qian}}, \bibinfo {author} {\bibfnamefont {L.}~\bibnamefont {Zhu}}, \bibinfo
  {author} {\bibfnamefont {K.~H.}\ \bibnamefont {Ahn}}, \ and\ \bibinfo
  {author} {\bibfnamefont {C.}~\bibnamefont {Prodan}},\ }\href {\doibase
  10.1103/PhysRevLett.125.225501} {\bibfield  {journal} {\bibinfo  {journal}
  {Phys. Rev. Lett.}\ }\textbf {\bibinfo {volume} {125}},\ \bibinfo {pages}
  {225501} (\bibinfo {year} {2020})}\BibitemShut {NoStop}%
\bibitem [{\citenamefont {Slager}\ \emph {et~al.}(2012)\citenamefont {Slager},
  \citenamefont {Mesaros}, \citenamefont {Juri{\v{c}}i{\'{c}}},\ and\
  \citenamefont {Zaanen}}]{Slager2012}%
  \BibitemOpen
  \bibfield  {author} {\bibinfo {author} {\bibfnamefont {R.-J.}\ \bibnamefont
  {Slager}}, \bibinfo {author} {\bibfnamefont {A.}~\bibnamefont {Mesaros}},
  \bibinfo {author} {\bibfnamefont {V.}~\bibnamefont {Juri{\v{c}}i{\'{c}}}}, \
  and\ \bibinfo {author} {\bibfnamefont {J.}~\bibnamefont {Zaanen}},\ }\href
  {\doibase 10.1038/nphys2513} {\bibfield  {journal} {\bibinfo  {journal}
  {Nature Physics}\ }\textbf {\bibinfo {volume} {9}},\ \bibinfo {pages} {98}
  (\bibinfo {year} {2012})}\BibitemShut {NoStop}%
\bibitem [{\citenamefont {Fang}\ \emph {et~al.}(2012)\citenamefont {Fang},
  \citenamefont {Gilbert},\ and\ \citenamefont
  {Bernevig}}]{PhysRevB.86.115112}%
  \BibitemOpen
  \bibfield  {author} {\bibinfo {author} {\bibfnamefont {C.}~\bibnamefont
  {Fang}}, \bibinfo {author} {\bibfnamefont {M.~J.}\ \bibnamefont {Gilbert}}, \
  and\ \bibinfo {author} {\bibfnamefont {B.~A.}\ \bibnamefont {Bernevig}},\
  }\href {\doibase 10.1103/PhysRevB.86.115112} {\bibfield  {journal} {\bibinfo
  {journal} {Phys. Rev. B}\ }\textbf {\bibinfo {volume} {86}},\ \bibinfo
  {pages} {115112} (\bibinfo {year} {2012})}\BibitemShut {NoStop}%
\bibitem [{\citenamefont {Kruthoff}\ \emph {et~al.}(2017)\citenamefont
  {Kruthoff}, \citenamefont {de~Boer}, \citenamefont {van Wezel}, \citenamefont
  {Kane},\ and\ \citenamefont {Slager}}]{PhysRevX.7.041069}%
  \BibitemOpen
  \bibfield  {author} {\bibinfo {author} {\bibfnamefont {J.}~\bibnamefont
  {Kruthoff}}, \bibinfo {author} {\bibfnamefont {J.}~\bibnamefont {de~Boer}},
  \bibinfo {author} {\bibfnamefont {J.}~\bibnamefont {van Wezel}}, \bibinfo
  {author} {\bibfnamefont {C.~L.}\ \bibnamefont {Kane}}, \ and\ \bibinfo
  {author} {\bibfnamefont {R.-J.}\ \bibnamefont {Slager}},\ }\href {\doibase
  10.1103/PhysRevX.7.041069} {\bibfield  {journal} {\bibinfo  {journal} {Phys.
  Rev. X}\ }\textbf {\bibinfo {volume} {7}},\ \bibinfo {pages} {041069}
  (\bibinfo {year} {2017})}\BibitemShut {NoStop}%
\bibitem [{\citenamefont {Bradlyn}\ \emph {et~al.}(2017)\citenamefont
  {Bradlyn}, \citenamefont {Elcoro}, \citenamefont {Cano}, \citenamefont
  {Vergniory}, \citenamefont {Wang}, \citenamefont {Felser}, \citenamefont
  {Aroyo},\ and\ \citenamefont {Bernevig}}]{Bradlyn2017}%
  \BibitemOpen
  \bibfield  {author} {\bibinfo {author} {\bibfnamefont {B.}~\bibnamefont
  {Bradlyn}}, \bibinfo {author} {\bibfnamefont {L.}~\bibnamefont {Elcoro}},
  \bibinfo {author} {\bibfnamefont {J.}~\bibnamefont {Cano}}, \bibinfo {author}
  {\bibfnamefont {M.~G.}\ \bibnamefont {Vergniory}}, \bibinfo {author}
  {\bibfnamefont {Z.}~\bibnamefont {Wang}}, \bibinfo {author} {\bibfnamefont
  {C.}~\bibnamefont {Felser}}, \bibinfo {author} {\bibfnamefont {M.~I.}\
  \bibnamefont {Aroyo}}, \ and\ \bibinfo {author} {\bibfnamefont {B.~A.}\
  \bibnamefont {Bernevig}},\ }\href {\doibase 10.1038/nature23268} {\bibfield
  {journal} {\bibinfo  {journal} {Nature}\ }\textbf {\bibinfo {volume} {547}},\
  \bibinfo {pages} {298} (\bibinfo {year} {2017})}\BibitemShut {NoStop}%
\bibitem [{\citenamefont {Po}\ \emph {et~al.}(2017)\citenamefont {Po},
  \citenamefont {Vishwanath},\ and\ \citenamefont {Watanabe}}]{Po17p1}%
  \BibitemOpen
  \bibfield  {author} {\bibinfo {author} {\bibfnamefont {H.~C.}\ \bibnamefont
  {Po}}, \bibinfo {author} {\bibfnamefont {A.}~\bibnamefont {Vishwanath}}, \
  and\ \bibinfo {author} {\bibfnamefont {H.}~\bibnamefont {Watanabe}},\ }\href
  {\doibase 10.1038/s41467-017-00724-z} {\bibfield  {journal} {\bibinfo
  {journal} {Nature Communications}\ }\textbf {\bibinfo {volume} {8}} (\bibinfo
  {year} {2017}),\ 10.1038/s41467-017-00724-z}\BibitemShut {NoStop}%
\bibitem [{\citenamefont {Khalaf}\ \emph {et~al.}(2018)\citenamefont {Khalaf},
  \citenamefont {Po}, \citenamefont {Vishwanath},\ and\ \citenamefont
  {Watanabe}}]{PhysRevX.8.031070}%
  \BibitemOpen
  \bibfield  {author} {\bibinfo {author} {\bibfnamefont {E.}~\bibnamefont
  {Khalaf}}, \bibinfo {author} {\bibfnamefont {H.~C.}\ \bibnamefont {Po}},
  \bibinfo {author} {\bibfnamefont {A.}~\bibnamefont {Vishwanath}}, \ and\
  \bibinfo {author} {\bibfnamefont {H.}~\bibnamefont {Watanabe}},\ }\href
  {\doibase 10.1103/PhysRevX.8.031070} {\bibfield  {journal} {\bibinfo
  {journal} {Phys. Rev. X}\ }\textbf {\bibinfo {volume} {8}},\ \bibinfo {pages}
  {031070} (\bibinfo {year} {2018})}\BibitemShut {NoStop}%
\bibitem [{\citenamefont {Ono}\ and\ \citenamefont
  {Watanabe}(2018)}]{PhysRevB.98.115150}%
  \BibitemOpen
  \bibfield  {author} {\bibinfo {author} {\bibfnamefont {S.}~\bibnamefont
  {Ono}}\ and\ \bibinfo {author} {\bibfnamefont {H.}~\bibnamefont {Watanabe}},\
  }\href {\doibase 10.1103/PhysRevB.98.115150} {\bibfield  {journal} {\bibinfo
  {journal} {Phys. Rev. B}\ }\textbf {\bibinfo {volume} {98}},\ \bibinfo
  {pages} {115150} (\bibinfo {year} {2018})}\BibitemShut {NoStop}%
\bibitem [{\citenamefont {Bouhon}\ \emph {et~al.}(2019)\citenamefont {Bouhon},
  \citenamefont {Black-Schaffer},\ and\ \citenamefont
  {Slager}}]{PhysRevB.100.195135}%
  \BibitemOpen
  \bibfield  {author} {\bibinfo {author} {\bibfnamefont {A.}~\bibnamefont
  {Bouhon}}, \bibinfo {author} {\bibfnamefont {A.~M.}\ \bibnamefont
  {Black-Schaffer}}, \ and\ \bibinfo {author} {\bibfnamefont {R.-J.}\
  \bibnamefont {Slager}},\ }\href {\doibase 10.1103/PhysRevB.100.195135}
  {\bibfield  {journal} {\bibinfo  {journal} {Phys. Rev. B}\ }\textbf {\bibinfo
  {volume} {100}},\ \bibinfo {pages} {195135(R)} (\bibinfo {year}
  {2019})}\BibitemShut {NoStop}%
\bibitem [{\citenamefont {Tang}\ \emph
  {et~al.}(2019{\natexlab{a}})\citenamefont {Tang}, \citenamefont {Po},
  \citenamefont {Vishwanath},\ and\ \citenamefont {Wan}}]{Tang19peaau8725}%
  \BibitemOpen
  \bibfield  {author} {\bibinfo {author} {\bibfnamefont {F.}~\bibnamefont
  {Tang}}, \bibinfo {author} {\bibfnamefont {H.~C.}\ \bibnamefont {Po}},
  \bibinfo {author} {\bibfnamefont {A.}~\bibnamefont {Vishwanath}}, \ and\
  \bibinfo {author} {\bibfnamefont {X.}~\bibnamefont {Wan}},\ }\href {\doibase
  10.1126/sciadv.aau8725} {\bibfield  {journal} {\bibinfo  {journal} {Science
  Advances}\ }\textbf {\bibinfo {volume} {5}},\ \bibinfo {pages} {eaau8725}
  (\bibinfo {year} {2019}{\natexlab{a}})}\BibitemShut {NoStop}%
\bibitem [{\citenamefont {Tang}\ \emph
  {et~al.}(2019{\natexlab{b}})\citenamefont {Tang}, \citenamefont {Po},
  \citenamefont {Vishwanath},\ and\ \citenamefont {Wan}}]{Tang19p470}%
  \BibitemOpen
  \bibfield  {author} {\bibinfo {author} {\bibfnamefont {F.}~\bibnamefont
  {Tang}}, \bibinfo {author} {\bibfnamefont {H.~C.}\ \bibnamefont {Po}},
  \bibinfo {author} {\bibfnamefont {A.}~\bibnamefont {Vishwanath}}, \ and\
  \bibinfo {author} {\bibfnamefont {X.}~\bibnamefont {Wan}},\ }\href {\doibase
  10.1038/s41567-019-0418-7} {\bibfield  {journal} {\bibinfo  {journal} {Nature
  Physics}\ }\textbf {\bibinfo {volume} {15}},\ \bibinfo {pages} {470}
  (\bibinfo {year} {2019}{\natexlab{b}})}\BibitemShut {NoStop}%
\bibitem [{\citenamefont {Tang}\ \emph
  {et~al.}(2019{\natexlab{c}})\citenamefont {Tang}, \citenamefont {Po},
  \citenamefont {Vishwanath},\ and\ \citenamefont {Wan}}]{Tang19p486}%
  \BibitemOpen
  \bibfield  {author} {\bibinfo {author} {\bibfnamefont {F.}~\bibnamefont
  {Tang}}, \bibinfo {author} {\bibfnamefont {H.~C.}\ \bibnamefont {Po}},
  \bibinfo {author} {\bibfnamefont {A.}~\bibnamefont {Vishwanath}}, \ and\
  \bibinfo {author} {\bibfnamefont {X.}~\bibnamefont {Wan}},\ }\href {\doibase
  10.1038/s41586-019-0937-5} {\bibfield  {journal} {\bibinfo  {journal}
  {Nature}\ }\textbf {\bibinfo {volume} {566}},\ \bibinfo {pages} {486}
  (\bibinfo {year} {2019}{\natexlab{c}})}\BibitemShut {NoStop}%
\bibitem [{\citenamefont {Zhang}\ \emph {et~al.}(2019)\citenamefont {Zhang},
  \citenamefont {Jiang}, \citenamefont {Song}, \citenamefont {Huang},
  \citenamefont {He}, \citenamefont {Fang}, \citenamefont {Weng},\ and\
  \citenamefont {Fang}}]{Zhang19p475}%
  \BibitemOpen
  \bibfield  {author} {\bibinfo {author} {\bibfnamefont {T.}~\bibnamefont
  {Zhang}}, \bibinfo {author} {\bibfnamefont {Y.}~\bibnamefont {Jiang}},
  \bibinfo {author} {\bibfnamefont {Z.}~\bibnamefont {Song}}, \bibinfo {author}
  {\bibfnamefont {H.}~\bibnamefont {Huang}}, \bibinfo {author} {\bibfnamefont
  {Y.}~\bibnamefont {He}}, \bibinfo {author} {\bibfnamefont {Z.}~\bibnamefont
  {Fang}}, \bibinfo {author} {\bibfnamefont {H.}~\bibnamefont {Weng}}, \ and\
  \bibinfo {author} {\bibfnamefont {C.}~\bibnamefont {Fang}},\ }\href {\doibase
  10.1038/s41586-019-0944-6} {\bibfield  {journal} {\bibinfo  {journal}
  {Nature}\ }\textbf {\bibinfo {volume} {566}},\ \bibinfo {pages} {475}
  (\bibinfo {year} {2019})}\BibitemShut {NoStop}%
\bibitem [{\citenamefont {Vergniory}\ \emph {et~al.}(2019)\citenamefont
  {Vergniory}, \citenamefont {Elcoro}, \citenamefont {Felser}, \citenamefont
  {Regnault}, \citenamefont {Bernevig},\ and\ \citenamefont
  {Wang}}]{Vergniory2019}%
  \BibitemOpen
  \bibfield  {author} {\bibinfo {author} {\bibfnamefont {M.~G.}\ \bibnamefont
  {Vergniory}}, \bibinfo {author} {\bibfnamefont {L.}~\bibnamefont {Elcoro}},
  \bibinfo {author} {\bibfnamefont {C.}~\bibnamefont {Felser}}, \bibinfo
  {author} {\bibfnamefont {N.}~\bibnamefont {Regnault}}, \bibinfo {author}
  {\bibfnamefont {B.~A.}\ \bibnamefont {Bernevig}}, \ and\ \bibinfo {author}
  {\bibfnamefont {Z.}~\bibnamefont {Wang}},\ }\href {\doibase
  10.1038/s41586-019-0954-4} {\bibfield  {journal} {\bibinfo  {journal}
  {Nature}\ }\textbf {\bibinfo {volume} {566}},\ \bibinfo {pages} {480}
  (\bibinfo {year} {2019})}\BibitemShut {NoStop}%
\bibitem [{\citenamefont {Vergniory}\ \emph {et~al.}(2021)\citenamefont
  {Vergniory}, \citenamefont {Wieder}, \citenamefont {Elcoro}, \citenamefont
  {Parkin}, \citenamefont {Felser}, \citenamefont {Bernevig},\ and\
  \citenamefont {Regnault}}]{Maia21p2105.09954}%
  \BibitemOpen
  \bibfield  {author} {\bibinfo {author} {\bibfnamefont {M.~G.}\ \bibnamefont
  {Vergniory}}, \bibinfo {author} {\bibfnamefont {B.~J.}\ \bibnamefont
  {Wieder}}, \bibinfo {author} {\bibfnamefont {L.}~\bibnamefont {Elcoro}},
  \bibinfo {author} {\bibfnamefont {S.~S.~P.}\ \bibnamefont {Parkin}}, \bibinfo
  {author} {\bibfnamefont {C.}~\bibnamefont {Felser}}, \bibinfo {author}
  {\bibfnamefont {B.~A.}\ \bibnamefont {Bernevig}}, \ and\ \bibinfo {author}
  {\bibfnamefont {N.}~\bibnamefont {Regnault}},\ }\href@noop {} {\enquote
  {\bibinfo {title} {All topological bands of all stoichiometric materials},}\
  } (\bibinfo {year} {2021}),\ \Eprint {http://arxiv.org/abs/arXiv:2105.09954}
  {arXiv:2105.09954} \BibitemShut {NoStop}%
\bibitem [{\citenamefont {Fu}\ and\ \citenamefont {Kane}(2007)}]{Fu07p045302}%
  \BibitemOpen
  \bibfield  {author} {\bibinfo {author} {\bibfnamefont {L.}~\bibnamefont
  {Fu}}\ and\ \bibinfo {author} {\bibfnamefont {C.~L.}\ \bibnamefont {Kane}},\
  }\href {\doibase 10.1103/PhysRevB.76.045302} {\bibfield  {journal} {\bibinfo
  {journal} {Phys. Rev. B}\ }\textbf {\bibinfo {volume} {76}},\ \bibinfo
  {pages} {045302} (\bibinfo {year} {2007})}\BibitemShut {NoStop}%
\bibitem [{\citenamefont {Hasan}\ and\ \citenamefont
  {Moore}(2011)}]{Hasan11p55}%
  \BibitemOpen
  \bibfield  {author} {\bibinfo {author} {\bibfnamefont {M.~Z.}\ \bibnamefont
  {Hasan}}\ and\ \bibinfo {author} {\bibfnamefont {J.~E.}\ \bibnamefont
  {Moore}},\ }\href {\doibase 10.1146/annurev-conmatphys-062910-140432}
  {\bibfield  {journal} {\bibinfo  {journal} {Annual Review of Condensed Matter
  Physics}\ }\textbf {\bibinfo {volume} {2}},\ \bibinfo {pages} {55} (\bibinfo
  {year} {2011})},\ \Eprint
  {http://arxiv.org/abs/https://doi.org/10.1146/annurev-conmatphys-062910-140432}
  {https://doi.org/10.1146/annurev-conmatphys-062910-140432} \BibitemShut
  {NoStop}%
\bibitem [{\citenamefont {Hsieh}\ \emph {et~al.}(2008)\citenamefont {Hsieh},
  \citenamefont {Qian}, \citenamefont {Wray}, \citenamefont {Xia},
  \citenamefont {Hor}, \citenamefont {Cava},\ and\ \citenamefont
  {Hasan}}]{Hsieh08p970}%
  \BibitemOpen
  \bibfield  {author} {\bibinfo {author} {\bibfnamefont {D.}~\bibnamefont
  {Hsieh}}, \bibinfo {author} {\bibfnamefont {D.}~\bibnamefont {Qian}},
  \bibinfo {author} {\bibfnamefont {L.}~\bibnamefont {Wray}}, \bibinfo {author}
  {\bibfnamefont {Y.}~\bibnamefont {Xia}}, \bibinfo {author} {\bibfnamefont
  {Y.~S.}\ \bibnamefont {Hor}}, \bibinfo {author} {\bibfnamefont {R.~J.}\
  \bibnamefont {Cava}}, \ and\ \bibinfo {author} {\bibfnamefont {M.~Z.}\
  \bibnamefont {Hasan}},\ }\href {\doibase 10.1038/nature06843} {\bibfield
  {journal} {\bibinfo  {journal} {Nature}\ }\textbf {\bibinfo {volume} {452}},\
  \bibinfo {pages} {970} (\bibinfo {year} {2008})}\BibitemShut {NoStop}%
\bibitem [{\citenamefont {Xia}\ \emph {et~al.}(2009)\citenamefont {Xia},
  \citenamefont {Qian}, \citenamefont {Hsieh}, \citenamefont {Wray},
  \citenamefont {Pal}, \citenamefont {Lin}, \citenamefont {Bansil},
  \citenamefont {Grauer}, \citenamefont {Hor}, \citenamefont {Cava},\ and\
  \citenamefont {Hasan}}]{Xia09p398}%
  \BibitemOpen
  \bibfield  {author} {\bibinfo {author} {\bibfnamefont {Y.}~\bibnamefont
  {Xia}}, \bibinfo {author} {\bibfnamefont {D.}~\bibnamefont {Qian}}, \bibinfo
  {author} {\bibfnamefont {D.}~\bibnamefont {Hsieh}}, \bibinfo {author}
  {\bibfnamefont {L.}~\bibnamefont {Wray}}, \bibinfo {author} {\bibfnamefont
  {A.}~\bibnamefont {Pal}}, \bibinfo {author} {\bibfnamefont {H.}~\bibnamefont
  {Lin}}, \bibinfo {author} {\bibfnamefont {A.}~\bibnamefont {Bansil}},
  \bibinfo {author} {\bibfnamefont {D.}~\bibnamefont {Grauer}}, \bibinfo
  {author} {\bibfnamefont {Y.~S.}\ \bibnamefont {Hor}}, \bibinfo {author}
  {\bibfnamefont {R.~J.}\ \bibnamefont {Cava}}, \ and\ \bibinfo {author}
  {\bibfnamefont {M.~Z.}\ \bibnamefont {Hasan}},\ }\href {\doibase
  10.1038/nphys1274} {\bibfield  {journal} {\bibinfo  {journal} {Nature
  Physics}\ }\textbf {\bibinfo {volume} {5}},\ \bibinfo {pages} {398} (\bibinfo
  {year} {2009})}\BibitemShut {NoStop}%
\bibitem [{\citenamefont {Zhang}\ \emph {et~al.}(2009)\citenamefont {Zhang},
  \citenamefont {Qin}, \citenamefont {Teng}, \citenamefont {Guo}, \citenamefont
  {Guo}, \citenamefont {Dai}, \citenamefont {Fang},\ and\ \citenamefont
  {Wu}}]{Zhang09p053114}%
  \BibitemOpen
  \bibfield  {author} {\bibinfo {author} {\bibfnamefont {G.}~\bibnamefont
  {Zhang}}, \bibinfo {author} {\bibfnamefont {H.}~\bibnamefont {Qin}}, \bibinfo
  {author} {\bibfnamefont {J.}~\bibnamefont {Teng}}, \bibinfo {author}
  {\bibfnamefont {J.}~\bibnamefont {Guo}}, \bibinfo {author} {\bibfnamefont
  {Q.}~\bibnamefont {Guo}}, \bibinfo {author} {\bibfnamefont {X.}~\bibnamefont
  {Dai}}, \bibinfo {author} {\bibfnamefont {Z.}~\bibnamefont {Fang}}, \ and\
  \bibinfo {author} {\bibfnamefont {K.}~\bibnamefont {Wu}},\ }\href {\doibase
  10.1063/1.3200237} {\bibfield  {journal} {\bibinfo  {journal} {Applied
  Physics Letters}\ }\textbf {\bibinfo {volume} {95}},\ \bibinfo {pages}
  {053114} (\bibinfo {year} {2009})}\BibitemShut {NoStop}%
\bibitem [{\citenamefont {Benalcazar}\ \emph {et~al.}(2017)\citenamefont
  {Benalcazar}, \citenamefont {Bernevig},\ and\ \citenamefont
  {Hughes}}]{Benalcazar2017}%
  \BibitemOpen
  \bibfield  {author} {\bibinfo {author} {\bibfnamefont {W.~A.}\ \bibnamefont
  {Benalcazar}}, \bibinfo {author} {\bibfnamefont {B.~A.}\ \bibnamefont
  {Bernevig}}, \ and\ \bibinfo {author} {\bibfnamefont {T.~L.}\ \bibnamefont
  {Hughes}},\ }\href {\doibase 10.1126/science.aah6442} {\bibfield  {journal}
  {\bibinfo  {journal} {Science}\ }\textbf {\bibinfo {volume} {357}},\ \bibinfo
  {pages} {61} (\bibinfo {year} {2017})}\BibitemShut {NoStop}%
\bibitem [{\citenamefont {Langbehn}\ \emph {et~al.}(2017)\citenamefont
  {Langbehn}, \citenamefont {Peng}, \citenamefont {Trifunovic}, \citenamefont
  {von Oppen},\ and\ \citenamefont {Brouwer}}]{PhysRevLett.119.246401}%
  \BibitemOpen
  \bibfield  {author} {\bibinfo {author} {\bibfnamefont {J.}~\bibnamefont
  {Langbehn}}, \bibinfo {author} {\bibfnamefont {Y.}~\bibnamefont {Peng}},
  \bibinfo {author} {\bibfnamefont {L.}~\bibnamefont {Trifunovic}}, \bibinfo
  {author} {\bibfnamefont {F.}~\bibnamefont {von Oppen}}, \ and\ \bibinfo
  {author} {\bibfnamefont {P.~W.}\ \bibnamefont {Brouwer}},\ }\href {\doibase
  10.1103/PhysRevLett.119.246401} {\bibfield  {journal} {\bibinfo  {journal}
  {Phys. Rev. Lett.}\ }\textbf {\bibinfo {volume} {119}},\ \bibinfo {pages}
  {246401} (\bibinfo {year} {2017})}\BibitemShut {NoStop}%
\bibitem [{\citenamefont {Song}\ \emph {et~al.}(2017)\citenamefont {Song},
  \citenamefont {Fang},\ and\ \citenamefont {Fang}}]{PhysRevLett.119.246402}%
  \BibitemOpen
  \bibfield  {author} {\bibinfo {author} {\bibfnamefont {Z.}~\bibnamefont
  {Song}}, \bibinfo {author} {\bibfnamefont {Z.}~\bibnamefont {Fang}}, \ and\
  \bibinfo {author} {\bibfnamefont {C.}~\bibnamefont {Fang}},\ }\href {\doibase
  10.1103/PhysRevLett.119.246402} {\bibfield  {journal} {\bibinfo  {journal}
  {Phys. Rev. Lett.}\ }\textbf {\bibinfo {volume} {119}},\ \bibinfo {pages}
  {246402} (\bibinfo {year} {2017})}\BibitemShut {NoStop}%
\bibitem [{\citenamefont {Schindler}\ \emph {et~al.}(2018)\citenamefont
  {Schindler}, \citenamefont {Cook}, \citenamefont {Vergniory}, \citenamefont
  {Wang}, \citenamefont {Parkin}, \citenamefont {Bernevig},\ and\ \citenamefont
  {Neupert}}]{Schindler2018}%
  \BibitemOpen
  \bibfield  {author} {\bibinfo {author} {\bibfnamefont {F.}~\bibnamefont
  {Schindler}}, \bibinfo {author} {\bibfnamefont {A.~M.}\ \bibnamefont {Cook}},
  \bibinfo {author} {\bibfnamefont {M.~G.}\ \bibnamefont {Vergniory}}, \bibinfo
  {author} {\bibfnamefont {Z.}~\bibnamefont {Wang}}, \bibinfo {author}
  {\bibfnamefont {S.~S.~P.}\ \bibnamefont {Parkin}}, \bibinfo {author}
  {\bibfnamefont {B.~A.}\ \bibnamefont {Bernevig}}, \ and\ \bibinfo {author}
  {\bibfnamefont {T.}~\bibnamefont {Neupert}},\ }\href {\doibase
  10.1126/sciadv.aat0346} {\bibfield  {journal} {\bibinfo  {journal} {Science
  Advances}\ }\textbf {\bibinfo {volume} {4}},\ \bibinfo {pages} {eaat0346}
  (\bibinfo {year} {2018})}\BibitemShut {NoStop}%
\bibitem [{\citenamefont {Song}\ \emph {et~al.}(2020)\citenamefont {Song},
  \citenamefont {Elcoro}, \citenamefont {Xu}, \citenamefont {Regnault},\ and\
  \citenamefont {Bernevig}}]{PhysRevX.10.031001}%
  \BibitemOpen
  \bibfield  {author} {\bibinfo {author} {\bibfnamefont {Z.-D.}\ \bibnamefont
  {Song}}, \bibinfo {author} {\bibfnamefont {L.}~\bibnamefont {Elcoro}},
  \bibinfo {author} {\bibfnamefont {Y.-F.}\ \bibnamefont {Xu}}, \bibinfo
  {author} {\bibfnamefont {N.}~\bibnamefont {Regnault}}, \ and\ \bibinfo
  {author} {\bibfnamefont {B.~A.}\ \bibnamefont {Bernevig}},\ }\href {\doibase
  10.1103/PhysRevX.10.031001} {\bibfield  {journal} {\bibinfo  {journal} {Phys.
  Rev. X}\ }\textbf {\bibinfo {volume} {10}},\ \bibinfo {pages} {031001}
  (\bibinfo {year} {2020})}\BibitemShut {NoStop}%
\bibitem [{\citenamefont {Ahn}\ \emph {et~al.}(2018)\citenamefont {Ahn},
  \citenamefont {Kim}, \citenamefont {Kim},\ and\ \citenamefont
  {Yang}}]{Ahn18p106403}%
  \BibitemOpen
  \bibfield  {author} {\bibinfo {author} {\bibfnamefont {J.}~\bibnamefont
  {Ahn}}, \bibinfo {author} {\bibfnamefont {D.}~\bibnamefont {Kim}}, \bibinfo
  {author} {\bibfnamefont {Y.}~\bibnamefont {Kim}}, \ and\ \bibinfo {author}
  {\bibfnamefont {B.-J.}\ \bibnamefont {Yang}},\ }\href {\doibase
  10.1103/PhysRevLett.121.106403} {\bibfield  {journal} {\bibinfo  {journal}
  {Phys. Rev. Lett.}\ }\textbf {\bibinfo {volume} {121}},\ \bibinfo {pages}
  {106403} (\bibinfo {year} {2018})}\BibitemShut {NoStop}%
\bibitem [{\citenamefont {Wang}\ \emph {et~al.}(2019)\citenamefont {Wang},
  \citenamefont {Wieder}, \citenamefont {Li}, \citenamefont {Yan},\ and\
  \citenamefont {Bernevig}}]{Zhijun19p186401}%
  \BibitemOpen
  \bibfield  {author} {\bibinfo {author} {\bibfnamefont {Z.}~\bibnamefont
  {Wang}}, \bibinfo {author} {\bibfnamefont {B.~J.}\ \bibnamefont {Wieder}},
  \bibinfo {author} {\bibfnamefont {J.}~\bibnamefont {Li}}, \bibinfo {author}
  {\bibfnamefont {B.}~\bibnamefont {Yan}}, \ and\ \bibinfo {author}
  {\bibfnamefont {B.~A.}\ \bibnamefont {Bernevig}},\ }\href {\doibase
  10.1103/PhysRevLett.123.186401} {\bibfield  {journal} {\bibinfo  {journal}
  {Phys. Rev. Lett.}\ }\textbf {\bibinfo {volume} {123}},\ \bibinfo {pages}
  {186401} (\bibinfo {year} {2019})}\BibitemShut {NoStop}%
\bibitem [{\citenamefont {Park}\ \emph {et~al.}(2019)\citenamefont {Park},
  \citenamefont {Kim}, \citenamefont {Cho},\ and\ \citenamefont
  {Lee}}]{Park19p216803}%
  \BibitemOpen
  \bibfield  {author} {\bibinfo {author} {\bibfnamefont {M.~J.}\ \bibnamefont
  {Park}}, \bibinfo {author} {\bibfnamefont {Y.}~\bibnamefont {Kim}}, \bibinfo
  {author} {\bibfnamefont {G.~Y.}\ \bibnamefont {Cho}}, \ and\ \bibinfo
  {author} {\bibfnamefont {S.}~\bibnamefont {Lee}},\ }\href {\doibase
  10.1103/PhysRevLett.123.216803} {\bibfield  {journal} {\bibinfo  {journal}
  {Phys. Rev. Lett.}\ }\textbf {\bibinfo {volume} {123}},\ \bibinfo {pages}
  {216803} (\bibinfo {year} {2019})}\BibitemShut {NoStop}%
\bibitem [{\citenamefont {Lee}\ \emph {et~al.}(2020)\citenamefont {Lee},
  \citenamefont {Kim}, \citenamefont {Ahn},\ and\ \citenamefont
  {Yang}}]{Lee20p1}%
  \BibitemOpen
  \bibfield  {author} {\bibinfo {author} {\bibfnamefont {E.}~\bibnamefont
  {Lee}}, \bibinfo {author} {\bibfnamefont {R.}~\bibnamefont {Kim}}, \bibinfo
  {author} {\bibfnamefont {J.}~\bibnamefont {Ahn}}, \ and\ \bibinfo {author}
  {\bibfnamefont {B.-J.}\ \bibnamefont {Yang}},\ }\href {\doibase
  10.1038/s41535-019-0206-8} {\bibfield  {journal} {\bibinfo  {journal} {npj
  Quantum Materials}\ }\textbf {\bibinfo {volume} {5}} (\bibinfo {year}
  {2020}),\ 10.1038/s41535-019-0206-8}\BibitemShut {NoStop}%
\bibitem [{\citenamefont {Hughes}\ \emph {et~al.}(2011)\citenamefont {Hughes},
  \citenamefont {Prodan},\ and\ \citenamefont {Bernevig}}]{Hughes11p245132}%
  \BibitemOpen
  \bibfield  {author} {\bibinfo {author} {\bibfnamefont {T.~L.}\ \bibnamefont
  {Hughes}}, \bibinfo {author} {\bibfnamefont {E.}~\bibnamefont {Prodan}}, \
  and\ \bibinfo {author} {\bibfnamefont {B.~A.}\ \bibnamefont {Bernevig}},\
  }\href {\doibase 10.1103/PhysRevB.83.245132} {\bibfield  {journal} {\bibinfo
  {journal} {Phys. Rev. B}\ }\textbf {\bibinfo {volume} {83}},\ \bibinfo
  {pages} {245132} (\bibinfo {year} {2011})}\BibitemShut {NoStop}%
\bibitem [{\citenamefont {Alexandradinata}\ \emph {et~al.}(2014)\citenamefont
  {Alexandradinata}, \citenamefont {Dai},\ and\ \citenamefont
  {Bernevig}}]{Alexandradinata14p155114}%
  \BibitemOpen
  \bibfield  {author} {\bibinfo {author} {\bibfnamefont {A.}~\bibnamefont
  {Alexandradinata}}, \bibinfo {author} {\bibfnamefont {X.}~\bibnamefont
  {Dai}}, \ and\ \bibinfo {author} {\bibfnamefont {B.~A.}\ \bibnamefont
  {Bernevig}},\ }\href {\doibase 10.1103/PhysRevB.89.155114} {\bibfield
  {journal} {\bibinfo  {journal} {Phys. Rev. B}\ }\textbf {\bibinfo {volume}
  {89}},\ \bibinfo {pages} {155114} (\bibinfo {year} {2014})}\BibitemShut
  {NoStop}%
\bibitem [{\citenamefont {Ahn}\ \emph {et~al.}(2019)\citenamefont {Ahn},
  \citenamefont {Park}, \citenamefont {Kim}, \citenamefont {Kim},\ and\
  \citenamefont {Yang}}]{Ahn19p117101}%
  \BibitemOpen
  \bibfield  {author} {\bibinfo {author} {\bibfnamefont {J.}~\bibnamefont
  {Ahn}}, \bibinfo {author} {\bibfnamefont {S.}~\bibnamefont {Park}}, \bibinfo
  {author} {\bibfnamefont {D.}~\bibnamefont {Kim}}, \bibinfo {author}
  {\bibfnamefont {Y.}~\bibnamefont {Kim}}, \ and\ \bibinfo {author}
  {\bibfnamefont {B.-J.}\ \bibnamefont {Yang}},\ }\href {\doibase
  10.1088/1674-1056/ab4d3b} {\bibfield  {journal} {\bibinfo  {journal} {Chinese
  Physics B}\ }\textbf {\bibinfo {volume} {28}},\ \bibinfo {pages} {117101}
  (\bibinfo {year} {2019})}\BibitemShut {NoStop}%
\bibitem [{\citenamefont {Zak}(1989)}]{Zak89p2747}%
  \BibitemOpen
  \bibfield  {author} {\bibinfo {author} {\bibfnamefont {J.}~\bibnamefont
  {Zak}},\ }\href {\doibase 10.1103/PhysRevLett.62.2747} {\bibfield  {journal}
  {\bibinfo  {journal} {Phys. Rev. Lett.}\ }\textbf {\bibinfo {volume} {62}},\
  \bibinfo {pages} {2747} (\bibinfo {year} {1989})}\BibitemShut {NoStop}%
\bibitem [{\citenamefont {Kim}\ \emph {et~al.}(2015)\citenamefont {Kim},
  \citenamefont {Wieder}, \citenamefont {Kane},\ and\ \citenamefont
  {Rappe}}]{Kim15p036806}%
  \BibitemOpen
  \bibfield  {author} {\bibinfo {author} {\bibfnamefont {Y.}~\bibnamefont
  {Kim}}, \bibinfo {author} {\bibfnamefont {B.~J.}\ \bibnamefont {Wieder}},
  \bibinfo {author} {\bibfnamefont {C.~L.}\ \bibnamefont {Kane}}, \ and\
  \bibinfo {author} {\bibfnamefont {A.~M.}\ \bibnamefont {Rappe}},\ }\href
  {\doibase 10.1103/PhysRevLett.115.036806} {\bibfield  {journal} {\bibinfo
  {journal} {Phys. Rev. Lett.}\ }\textbf {\bibinfo {volume} {115}},\ \bibinfo
  {pages} {036806} (\bibinfo {year} {2015})}\BibitemShut {NoStop}%
\bibitem [{\citenamefont {Moore}\ and\ \citenamefont
  {Balents}(2007)}]{Moore07p121306}%
  \BibitemOpen
  \bibfield  {author} {\bibinfo {author} {\bibfnamefont {J.~E.}\ \bibnamefont
  {Moore}}\ and\ \bibinfo {author} {\bibfnamefont {L.}~\bibnamefont
  {Balents}},\ }\href {\doibase 10.1103/PhysRevB.75.121306} {\bibfield
  {journal} {\bibinfo  {journal} {Phys. Rev. B}\ }\textbf {\bibinfo {volume}
  {75}},\ \bibinfo {pages} {121306(R)} (\bibinfo {year} {2007})}\BibitemShut
  {NoStop}%
\bibitem [{\citenamefont {Teo}\ and\ \citenamefont
  {Kane}(2010)}]{Teo10p115120}%
  \BibitemOpen
  \bibfield  {author} {\bibinfo {author} {\bibfnamefont {J.~C.~Y.}\
  \bibnamefont {Teo}}\ and\ \bibinfo {author} {\bibfnamefont {C.~L.}\
  \bibnamefont {Kane}},\ }\href {\doibase 10.1103/PhysRevB.82.115120}
  {\bibfield  {journal} {\bibinfo  {journal} {Phys. Rev. B}\ }\textbf {\bibinfo
  {volume} {82}},\ \bibinfo {pages} {115120} (\bibinfo {year}
  {2010})}\BibitemShut {NoStop}%
\bibitem [{Note1()}]{Note1}%
  \BibitemOpen
  \bibinfo {note} {See the Supplemental Material for the detailed calculations
  methods.}\BibitemShut {Stop}%
\bibitem [{Note2()}]{Note2}%
  \BibitemOpen
  \bibinfo {note} {See the Supplemental Material for the detailed calculations
  methods and the list of the materials that realize the first
  2DSWI}\BibitemShut {NoStop}%
\bibitem [{Note3()}]{Note3}%
  \BibitemOpen
  \bibinfo {note} {See the Supplemental Material for the list of the
  materials.}\BibitemShut {Stop}%
\end{thebibliography}%
\end{document}